\documentclass[nohyperref]{article}

\usepackage{microtype}
\usepackage{graphicx}
\usepackage{subfigure}
\usepackage{booktabs} 

\usepackage{hyperref}
\usepackage{amsmath}
\usepackage{amssymb}
\usepackage{mathtools}
\usepackage{amsthm}
\usepackage{enumitem}
\usepackage{makecell}

\usepackage{amsmath,amsfonts,bm}

\newcommand{\set}[1]{\{#1\}}

\def\eqref#1{equation~\ref{#1}}

\def\1{\bm{1}}

\def\vc{{\bm{c}}}

\def\vf{{\bm{f}}}

\def\vh{{\bm{h}}}

\def\vn{{\bm{n}}}

\def\vq{{\bm{q}}}
\def\vr{{\bm{r}}}

\def\vt{{\bm{t}}}
\def\vu{{\bm{u}}}

\def\vx{{\bm{x}}}

\def\mD{{\bm{D}}}

\def\mO{{\bm{O}}}

\def\mR{{\bm{R}}}
\def\mS{{\bm{S}}}

\def\mU{{\bm{U}}}

\def\mX{{\bm{X}}}

\DeclareMathAlphabet{\mathsfit}{\encodingdefault}{\sfdefault}{m}{sl}
\SetMathAlphabet{\mathsfit}{bold}{\encodingdefault}{\sfdefault}{bx}{n}

\usepackage[accepted]{icml2023}

\usepackage{amsmath}
\usepackage{amssymb}
\usepackage{mathtools}
\usepackage{amsthm}

\usepackage[capitalize,noabbrev]{cleveref}

\theoremstyle{plain}

\theoremstyle{definition}

\theoremstyle{remark}

\usepackage[textsize=tiny]{todonotes}

\icmltitlerunning{Learning Subpocket Prototypes for Generalizable Structure-based Drug Design}

\begin{document}

\twocolumn[
\icmltitle{Learning Subpocket Prototypes for Generalizable Structure-based Drug Design}




\begin{icmlauthorlist}
\icmlauthor{Zaixi Zhang}{yyy,lab}
\icmlauthor{Qi Liu}{yyy,lab}
\end{icmlauthorlist}

\icmlaffiliation{yyy}{Anhui Province Key Lab of Big Data Analysis and Application, University of Science and Technology of China}
\icmlaffiliation{lab}{State Key Laboratory of Cognitive Intelligence}

\icmlcorrespondingauthor{Qi Liu}{qiliuql@ustc.edu.cn}

\icmlkeywords{Machine Learning, ICML}

\vskip 0.3in
]



\printAffiliationsAndNotice{} 

\begin{abstract}
Generating molecules with high binding affinities to target proteins (a.k.a. structure-based drug design) is a fundamental and challenging task in drug discovery. Recently, deep generative models have achieved remarkable success in generating 3D molecules conditioned on the protein pocket. 
However, most existing methods consider molecular generation for protein pockets independently while neglecting the underlying connections such as subpocket-level similarities. 
Subpockets are the local protein environments of ligand fragments and pockets with similar subpockets may bind the same molecular fragment (motif) even though their overall structures are different.
Therefore, the trained models can hardly generalize to unseen protein pockets in real-world applications.
In this paper, we propose a novel method DrugGPS for generalizable structure-based drug design. With the biochemical priors,
we propose to learn subpocket prototypes and construct a global interaction graph to model the interactions between subpocket prototypes and molecular motifs. 
Moreover, a hierarchical graph transformer encoder and motif-based 3D molecule generation scheme are used to improve the model's performance.  
The experimental results show
that our model consistently outperforms baselines in generating realistic drug candidates with high affinities in challenging out-of-distribution settings. 
\end{abstract}
\section{Introduction}
Structure-based drug design (SBDD), i.e., designing molecules with high affinities to target protein pockets is one critical and challenging task in drug discovery \cite{anderson2003process, blundell1996structure, verlinde1994structure, ferreira2015molecular, jin2020multi}. Traditionally this has been achieved with virtual screening that identifies candidate molecules from molecular databases based on rules such as molecular docking \cite{morris2008molecular, pagadala2017software} and molecular dynamics simulations \cite{hansson2002molecular, karplus2002molecular}. However, such exhaustive searches are time-consuming and infeasible to generate new molecules not existing in the database. Recently, a line of works leverage deep generative models to directly generate 3D molecules inside binding pockets \cite{luo2021autoregressive, liu2022generating, peng2022pocket2mol, zhang2023molecule}. 

\begin{figure}[t]
	\centering
	\subfigure[Aligned subpocket pairs]{\includegraphics[width=0.56\linewidth]{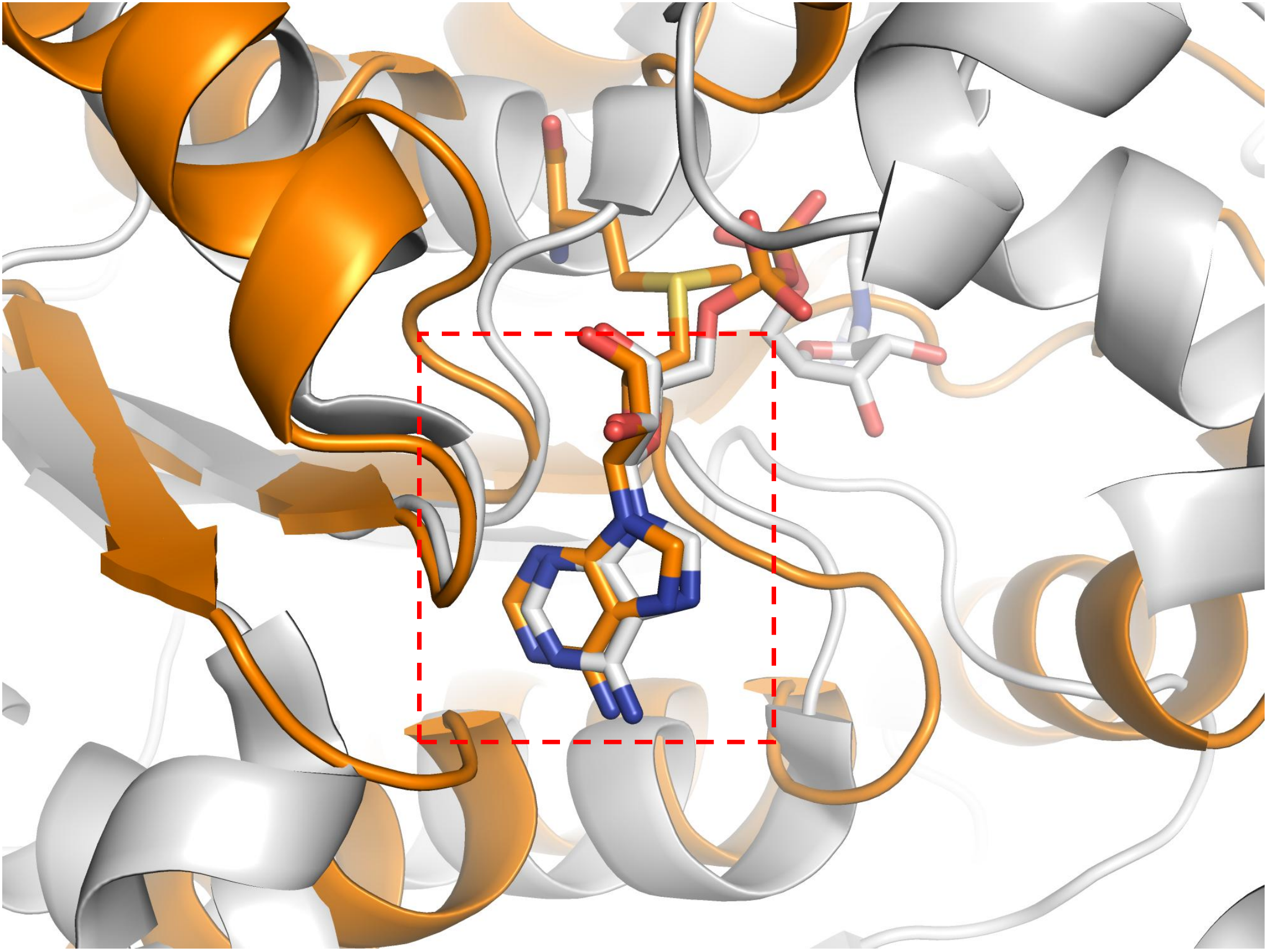}}
    \subfigure[Ligands]{\includegraphics[width=0.38\linewidth]{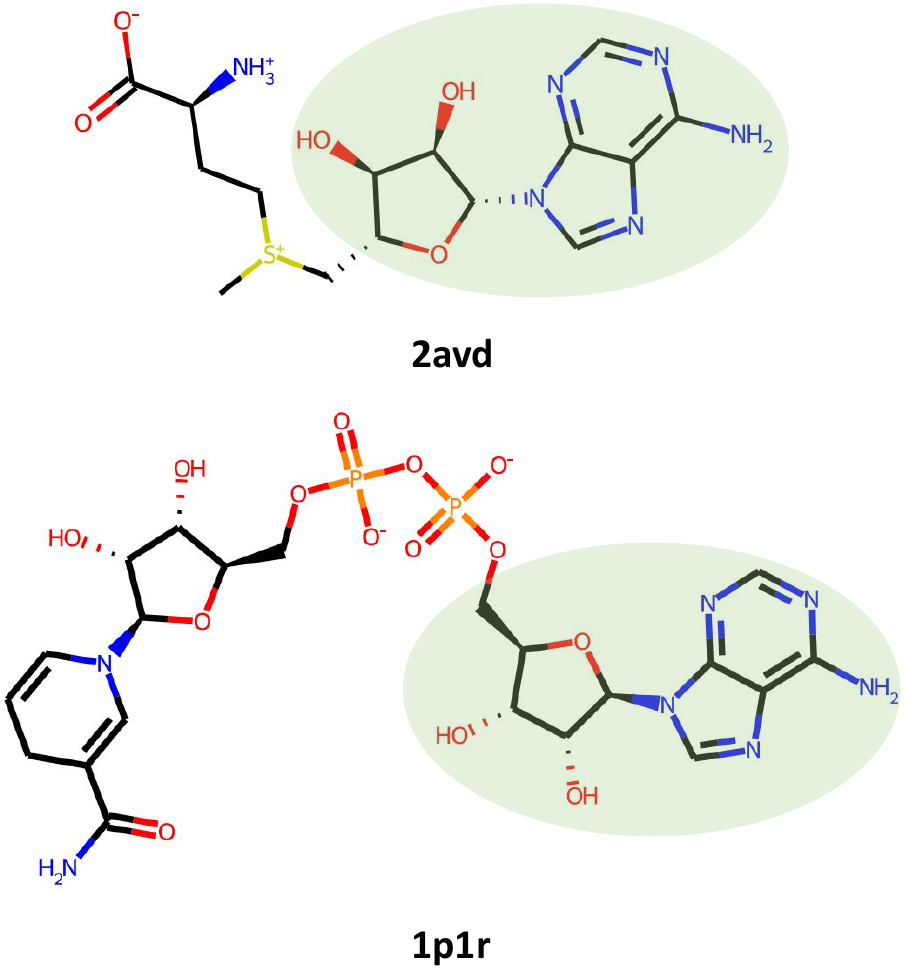}}
	    \vspace{-2mm}
	\caption{Illustration of our motivation. (a) Two proteins (PDB ID: 2avd and 1p1r) with low sequence similarity ($\le$10$\%$) have similar subpockets and bind to similar ligand fragments. 2avd is colored yellow and 1p1r is color white. The subpockets are aligned and highlighted with a red dashed box. (b) The molecular graphs of ligands binding to protein 2avd and 1p1r. Similar fragments in the subpockets are marked with green ovals. }
	\label{motivation}
\end{figure}
However, existing methods suffer from the generalization issue. The amount of high-quality protein-ligand complex data is rather limited and the target protein pocket may not be in the training dataset. In practice, when unpredictable events like COVID-19 occur, the generative models are required to generate molecules for new protein targets e.g., the main protease of SARS-CoV-2 \cite{zhang2020crystal}. Moreover, only atom-level interactions are considered and encoded in these works and the atom-by-atom generation may result in invalid molecules with unrealistic 3D structures.
More discussions are included in related works. 

In this paper, we propose \textbf{DrugGPS}, a structure-based \textbf{Drug} design method that is \textbf{G}eneralizable with \textbf{P}rotein \textbf{S}ubpocket prototypes to address the aforementioned challenges. Firstly, 
an atom-level graph and a residue-level graph are constructed to represent the binding context. A hierarchical 3D graph transformer is proposed to capture the hierarchical information. 
Secondly, to construct SBDD models that generalize well to unseen target protein pockets, we incorporate an effective biochemical prior into our model design:
\textbf{although two protein pockets might be dissimilar overall, they may still bind the same fragment if they share similar subpockets} \cite{kalliokoski2013subpocket}. Subpockets are defined as the local protein environment of the ligand fragments in protein-ligand complexes \cite{eguida2022target}. For example, in Figure. \ref{motivation}, two proteins (PDB ID: 2avd and 1p1r) with low sequence similarity ($\le$10$\%$) have similar subpockets and bind to similar ligand fragments. To capture the subpocket-level similarities/invariance among the binding pockets, we propose to learn subpocket prototypes and construct a global interaction graph to model the interactions between subpocket prototypes and molecular motifs (fragments) in the training process. 
To further highlight the subpocket-motif interactions, we employ an efficient binding analysis tool BINANA \cite{young2022binana} to identify polar contacts (hydrogen bonds).
In the generation process, the context representations are enriched with a global information fusion step and ligand molecules are generated motif-by-motif.  
In experiments, to mimic the real-world use case, we split the dataset
based on sequence-similarity and pocket-similarity and construct two out-of-distribution (OOD) settings. 
Experiment results
demonstrate that our method can generalize well to unseen pockets in the test set. The generated molecules not only show higher binding affinities and drug-likeness
but also contain more realistic substructures than the state-of-the-art baseline methods. Our key contributions include:
\begin{itemize}
\item In this paper, we propose \textbf{DrugGPS}, a structure-based \textbf{Drug} design method that is \textbf{G}eneralizable with \textbf{P}rotein \textbf{S}ubpocket prototypes.
\item A hierarchical 3D graph transformer is proposed to encode both the atom- and residue-level information.
\item We propose to construct the subpocket prototypes-molecular motif interaction graph in the training process. At the generation stage, molecules are generated motif-by-motif with the global interaction information. 
\item Experiments show that our model consistently outperforms baselines on generating realistic drug candidates with high affinities on challenging OOD settings.
\end{itemize}

\section{Related Works}
\subsection{Molecule Generation}
Recent years have witnessed the great success of deep generative models in molecule generation \cite{zhang2023equivariant, lee2022exploring, xie2021mars, yangknowledge}.
These models range from string-based \citep{gomez2018automatic} and graph-based methods \citep{jin2018junction, xie2021mars} to recent 3D geometry-based methods \citep{gebauer2019symmetry, luo2021autoregressive}. To enhance the validity of the generated molecules, some models adopt prior knowledge of molecular fragments, also known as motifs or rationales, as building blocks to generate and optimize molecules \cite{jin2018junction, jin2020hierarchical, xie2021mars}.
However, the generated molecules could
hardly fit and bind to given pockets in practice if the 3D conditional information, e.g., the shape and chemical properties of the protein pockets are neglected.

\subsection{Structure-based Drug Design.} Structure-based drug design (SBDD) aims to directly generate 3D molecules binding to target protein pockets. LiGAN \citep{ragoza2022generating} first uses 3D CNN to encode the protein-ligand structures and generate ligands by atom fitting and bond inference from the predicted atom densities.
Some follow-up works leverage graph neural networks to encode the context information and sample atoms auto-regressively \cite{luo2021autoregressive, liu2022generating, peng2022pocket2mol}. For example, GraphBP \cite{liu2022generating} adopts the framework of normalizing flow \citep{rezende2015variational} and constructs local coordinate systems to predict atom types and relative positions; Pocket2Mol \citep{peng2022pocket2mol} adopts the geometric vector perceptrons \citep{jing2021equivariant} and the vector-based neural network \citep{deng2021vector} as the context encoder. Some recent works also leverage fragment-based methods \cite{green2021deepfrag, powers2022fragment, zhang2023molecule} or pretrained models \cite{long2022zeroshot} to generate more realistic molecules. 
For example, \citep{powers2022fragment} expands a small molecule fragment into a larger drug-like molecule binding to a given protein pocket.
However, most existing methods suffer from the generalization concern in practice where only low-quality and deficient data is available. 
On the contrary, DrugGPS leverage the priors of protein subpockets to build generalizable models. 

\subsection{Generalizable Drug Discovery}
The ability to successfully apply previously acquired knowledge/data to new situations is vital to drug discovery \cite{2022arXiv220109637J, yang2022learning, zhang2021graph}. To this end, many works study drug discovery problems under
out-of-distribution settings.
For example, MoleOOD \cite{yang2022learning} builds generalizable molecule representation learning models against distribution shifts by learning invariant molecular substructure. PAR \cite{DBLP:journals/corr/abs-2107-07994} uses a meta-learning strategy for few-shot molecular property prediction.
MOOD \cite{lee2022exploring} designs an out-of-distribution molecule generation scheme with score-based diffusion to explore chemical space. 
However, these methods can hardly be applied to the more challenging conditional molecule generation task, i.e., structure-based drug design.

\begin{figure*}[t]
	\centering
	\includegraphics[width=0.98\linewidth]{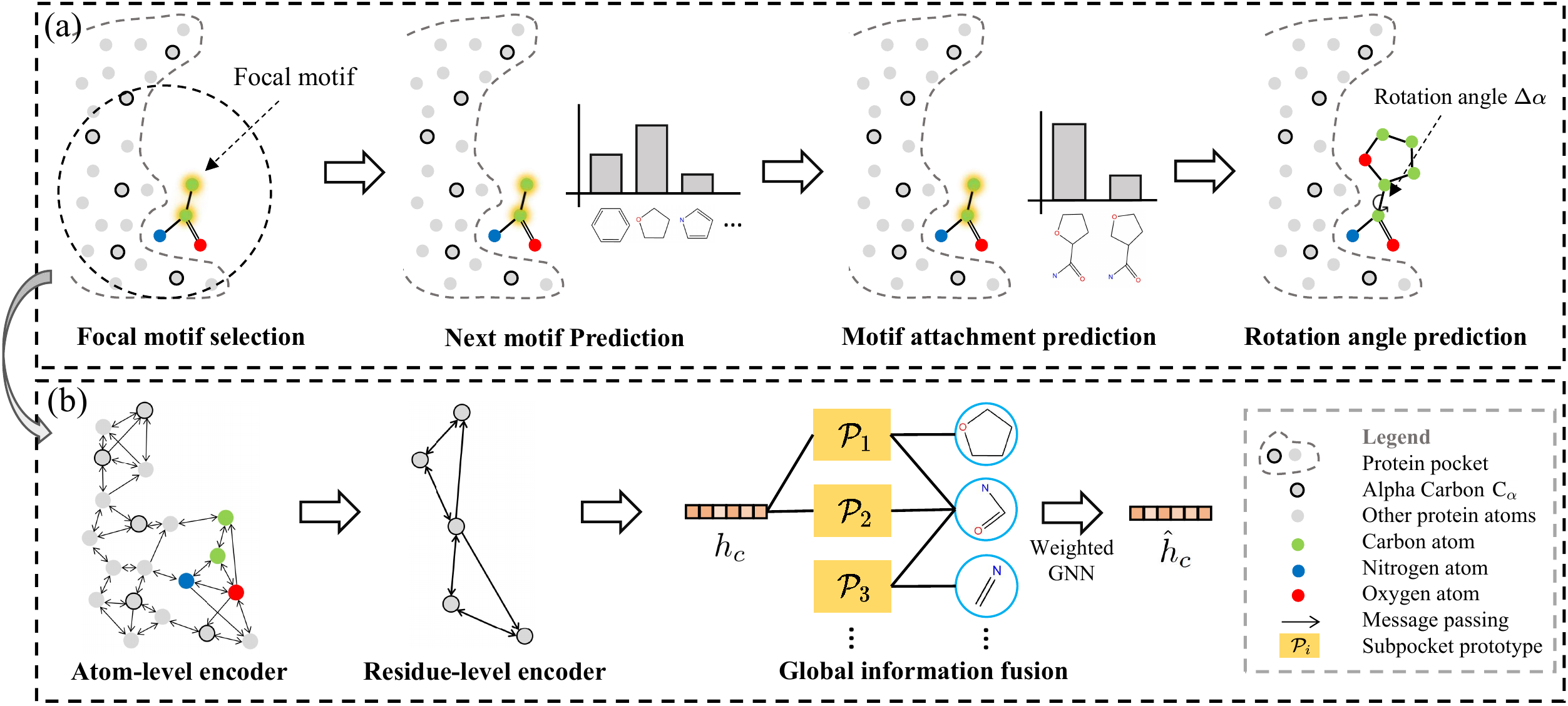}
	\caption{(a) The illustration of one generation step including four parts in our motif-based ligand generation scheme. (b) The hierarchical context encoder in DrugGPS. The global interaction information is further encoded into the subpocket embedding $h_c$ by a weighted GNN. }
	\label{illustration}
\end{figure*}

\section{Methods}
\subsection{Overview}
We first formalize the problem of structure-based drug design. Given a protein pocket-ligand complex, the 3D geometry of the ligand molecule can be represented as a set of atoms $\mathcal{G}^{mol}= \{(\bm{a}_i^{mol}, \bm{r}_i^{mol})\}$.
The protein pocket (i.e., binding site) can be similarly defined as $\mathcal{G}^{pro}= \{(\bm{a}_j^{pro}, \bm{r}_j^{pro})\}$.
In $\mathcal{G}^{mol}$ and $\mathcal{G}^{pro}$, $\bm{a}_i^{mol}$ and $\bm{a}_j^{pro}$ are one-hot vectors indicating the atom types and $\bm{r}_i^{mol}, \bm{r}_j^{pro} \in \mathbb{R}^3$ are the 3D cartesian coordinate vectors.
Formally, our objective is to learn a conditional generative model $p(\mathcal{G}^{mol}|\mathcal{G}^{pro})$ that captures the underlying dependencies of pocket-ligand pairs for 3D ligand molecule generation.

Specifically, we formulate the generation of molecules given binding pocket as a sequential decision process. Let $\phi$ be the generation model and $\mathcal{G}_t^{mol}$ be the intermediate molecule at the $t$-th step, the generation process is defined as follows:
\begin{align}
    &\mathcal{G}_t^{mol} = \phi(\mathcal{G}_{t-1}^{mol}, \mathcal{G}^{pro}), ~t>1\\
    &\mathcal{G}_1^{mol} = \phi(\mathcal{G}^{pro}), ~t=1.
\end{align}
Note that we generate molecules motif-by-motif, i.e., a set of atoms from the new motif are included into $\mathcal{G}_t^{mol}$ at each step. Figure. \ref{illustration}(a) demonstrates the four main parts in one
generation step, including (a) context encoding and focal motif selection, (b) next motif prediction, (c) motif attachment prediction, and (d) rotation angle prediction.

In this section, we first introduce the motif extraction procedure in Sec. \ref{motif extraction}. In Sec. \ref{encoder} and Sec. \ref{global interaction}, we will introduce the hierarchical context encoder and the construction of a global interaction graph, which are our main contributions to model architecture.  In Sec. \ref{generation} and Sec. \ref{training} we introduce the detailed generation procedures and derive the final training objectives. 
\subsection{Motif Vocabulary Construction}
\label{motif extraction}
Motif vocabulary construction aims to extract common molecular motifs from ligand molecules in whole dataset and construct a motif vocabulary $V_{\mathcal{M}} = \{\mathcal{M}_i\}$ for the follow-up molecule generation.
For the ease of motif extraction, molecules can be represented as 2D graphs $\mathcal{G}^{mol} = (\mathcal{V, E})$ with $\mathcal{V}$ as atoms set and $\mathcal{E}$ as covalent bonds set. Similarly, a motif $\mathcal{M}_i = (\mathcal{V}_i, \mathcal{E}_i)$ is defined as a molecular subgraph. Each molecule can also be represented as a set of motifs: $\mathcal{V} = \bigcup_i \mathcal{V}_i$ and $\mathcal{E} = \bigcup_i \mathcal{E}_i$.

\begin{figure}[t]
	\centering
	\includegraphics[width=0.98\linewidth]{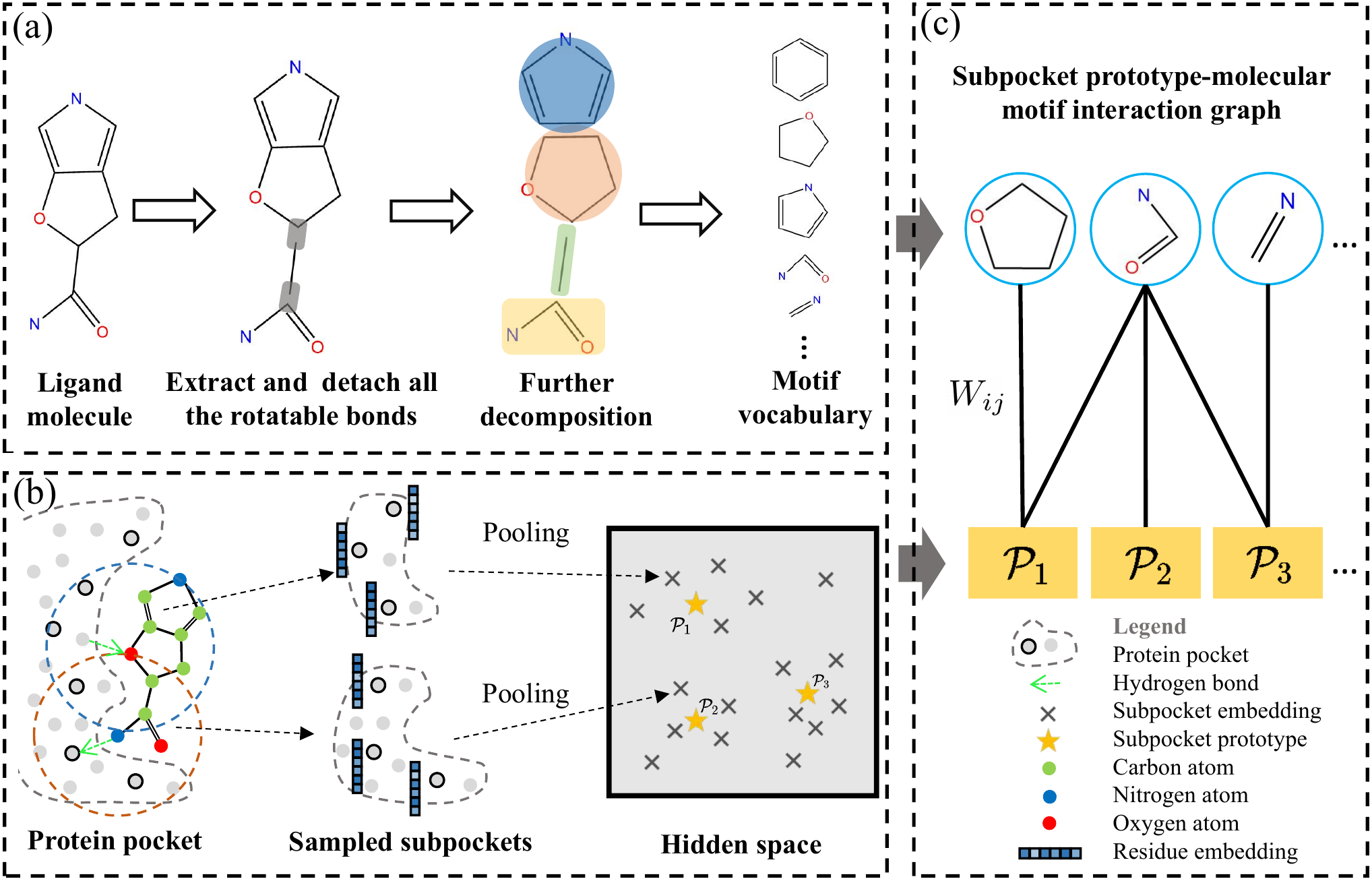}
	\caption{(a) The illustration of molecular motif extraction. (b) The sampled subpockets and subpocket prototypes. (c) The constructed Subpocket prototype-molecular motif interaction graph.}
	\label{meta graph}
\end{figure}

Figure. \ref{meta graph}(a) shows the procedures to fragment molecules and construct the motif vocabulary. To extract structural motifs, we first decompose a molecule $\mathcal{G}^{mol}$ into molecular substructures $\mathcal{G}_1, \cdots, \mathcal{G}_n$ by extracting and detaching all the rotatable bonds that will not violate the chemical validity. A bond in a molecule is rotatable if cutting this bond creates two connected components of the molecule, each of which has at least two atoms. We select $\mathcal{G}_i$ as a motif if its occurrence in the whole training set is more than $\tau$. We can select hyperparameter $\tau$ to control the size of the motif vocabulary $V_{\mathcal{M}}$ ranging from around 500 to over 2000. If $\mathcal{G}_i$ is not selected as a motif, we further decompose it into finer rings and bonds and select them as motifs.
As the bond length/angles in motifs are largely fixed, we employ RDkit \citep{bento2020open} to efficiently determine the 3D structures of motifs and trains neural networks to predict the torsion angles of rotatable bonds.


\subsection{Hierachical Context Encoder}
\label{encoder}
Inspired by the intrinsic hierarchical structure of protein \cite{stoker2015general}, we propose a hierarchical context encoder based on graph transformer \cite{min2022transformer, zhang2022hierarchical} to capture the context information of binding sites. Specifically, it includes an atom-level encoder and a residue-level encoder as described below.  
\subsubsection{Atom-level Encoder}
For the atom-level encoding, a context 3D graph $\mathcal{C}_{t-1}^a$ is first constructed by connecting the $K_a$ nearest neighboring atoms in $\mathcal{G}_{t-1}^{mol} \bigcup \mathcal{G}^{pro}$.
The atomic attributes are firstly mapped to  node embeddings ${\bm{h}^{(0)}_k}$ with a linear transformation layer. 
The edge embeddings $\bm{e}_{ij}$ are obtained by encoding pairwise distances with Gaussian functions \citep{schlichtkrull2018modeling}.
The 3D graph transformer consists of $L$ Transformer layers \cite{vaswani2017attention}. Each Transformer layer has two parts: a multi-head self-attention (MHA) module and a position-wise feed-forward network (FFN). Particularly, in the MHA module of the $l$-th layer $(1 \le l \le L)$, 
the queries are derived from the current node embeddings $\bm{h}_i^{(l)}$ while the keys and values from the relational information $\bm{r}_{ij}^{(l)} = \operatorname{Concat}(\bm{h}_j^{(l)}, \bm{e}_{ij}^{(l)})$ ($\operatorname{Concat}(\cdot)$ denotes concatenation) from neighboring nodes:
\begin{equation}
    \bm{q}_i^{(l)}= \mathbf{W}_{Q} \bm{h}_i^{(l)},~ \bm{k}_{ij}^{(l)} = \mathbf{W}_{K} \bm{r}_{ij}^{(l)},~ \bm{v}_{ij}^{(l)} = \mathbf{W}_{V} \bm{r}_{ij}^{(l)},
\end{equation}
where $\mathbf{W}_{Q}, \mathbf{W}_{K}$ and $\mathbf{W}_{V}$ are learnable transformation matrices.
Then, in each head $m \in \{1,2,\dots, M\}$ ($M$ is the total number of heads), the scaled dot-product attention mechanism is applied:
\begin{equation}
    \text{head}_i^m = \sum_{j\in\mathcal{N}(i)}\operatorname{Softmax}\left(\frac{{\bm{q}_i^{(l)}}^\top \cdot \bm{k}_{ij}^{(l)}}{\sqrt{d}}\right)\bm{v}_{ij}^{(l)},
\end{equation}
where $\mathcal{N}(i)$ denotes the neighbors of the $i$-th atom in $\mathcal{C}^a_{t-1}$ and $d$ is the dimension size of embeddings. Finally, the outputs from different heads are further concatenated and transformed to obtain the final output of MHA:
\begin{equation}
    \operatorname{MHA}_i=\operatorname{Concat}\left({\rm head}_{i}^1, \ldots, {\rm head}_{i}^M\right) \mathbf{W}_{O},
\end{equation}
where $\mathbf{W}_{O}$ is the output transformation matrix. The output of the atom-level encoder is a set of atom representations $\set{\vh_i}$. More architectural details are shown in Appendix. \ref{model architecture}.

\subsubsection{Residue-level Encoder}
The residue-level encoder only keeps the C$_\alpha$ atom of each residue and constructs a $K_r$ nearest neighbor graph $\mathcal{C}_{t-1}^{res}$ at the residue level. The $i$-th residue ($res_i$) can be represented by a feature vector $\vf_i$ describing its geometric and chemical characteristics including its dihedral angles, volume, polarity, charge, hydropathy, and hydrogen bond interactions. We concatenate the residue features with the sum of atom-level embeddings $\vh_k$ within that residue as the initial residue representation:
\begin{equation}
    \tilde{\vf}_i = {\rm Concat}\left(\vf_i, \sum_{k \in res_i}\nolimits \bm{h}_k\right).
\end{equation}
A local coordinate frame is built for each residue and the edge features $\bm{e}_{ij}^{res}$ between residues are computed describing the distance, direction, and orientation between neighboring residues~\citep{ingraham2019generative}. 
Lastly, the encoder takes the node and edge features into the residue-level graph transformer to compute the final representations of residues. The residue-level graph transformer architecture is similar to that of the atom-level encoder. More details of the residue-level encoder are shown in Appendix. \ref{model architecture}.

In summary, the output of our hierarchical encoder is a set of residue representations $\set{\vf_i}$ and atom representations $\set{\vh_i}$. Considering the distance range of pocket-ligand interactions \cite{marcou2007optimizing}, we sum all the residue representations within 6 {\AA} of the focal atom as the subpocket representations (Figure. \ref{meta graph}(b)). Since our encoder is based on the atom/residue attributes and pairwise relative distances, it is rotationally and translationally equivariant.

\subsection{Global Interaction Graph Construction}
\label{global interaction}
Most existing methods consider molecular generation for protein pockets independently while neglecting the underlying connections of subpocket-level similarities \cite{kalliokoski2013subpocket}. 
Here, we construct a global interaction graph to model the interactions between subpocket and ligand fragments in the whole dataset. As there are numerous subpockets in the dataset, we propose to cluster subpocket embeddings and derive representative subpocket prototypes (Figure. \ref{meta graph}(b)). Therefore, we have two kinds of nodes in the global interaction graph: subpocket prototype nodes and molecular motif nodes from the motif vocabulary (Figure. \ref{meta graph}(c)). 
The embeddings of subpocket prototypes and molecular motifs are dynamically updated during the training process.
We add an edge between a subpocket prototype node and a molecular motif node if a subpocket belonging to the cluster of the prototype binds to the motif in the training dataset. As the strength of the interaction is different, we calculate TF-IDF value as the edge weight $W_{ij}$ between a subpocket prototype $i$ and motif $j$:
\begin{equation}
    W_{ij} = C_{ij} \left({\rm log} \frac{1+N}{1+N_i}+1 \right),
    \label{edge weight}
\end{equation}
where $C_{ij}$ is the number of times that motif $j$ binds to subpockets belonging to prototype $i$, $N$ is the total number of subpocket prototypes, and $N_i$ is the number of subpocket prototypes binding to motif $i$. An edge has a larger weight if motif $j$ has a higher co-occurrence rate with prototype $i$ and binds to fewer other prototypes (higher specificity).

To further highlight the interactions between subpockets and molecular motifs, we employ an efficient binding analysis tool BINANA \cite{young2022binana} that is able to analyze the detailed interactions including hydrogen bonds, $\pi$-$\pi$ stacking, cation-$\pi$ interactions, electrostatic attraction, and hydrophobic with respect to each atom. When calculating edge weights $W_{ij}$ of the interaction graph, we only count subpocket-molecular motifs pairs with at least one hydrogen bond, which contributes much to binding affinity.

We update the subpocket prototypes on the fly during the model training process with an online K-Means algorithm. 
Specifically, to stabilize the training process, we update the prototypes i.e., centroids of clusters with momentum. Algorithm \ref{kmeans} shows the pseudo-codes of updating subpocket prototypes with a batch of input representations $\rm{H}$. Hyperparameter $\gamma$ is used to control the momentum. FindNearest($\cdot, \cdot$) denotes the function to find the nearest prototypes of inputs in Euclidean space and $S$ is the assignment matrix.  
Appendix. \ref{interaction graph} contains more details of the global interaction graph construction.

\begin{algorithm}[t]
\caption{Subpocket Prototypes Update Algorithm}
\label{kmeans}
\leftline{\textbf{Input}: subpocket representations $\rm{H}$, momentum hyper-} 
\leftline{parameter $\gamma$, Prototypes $\mathcal{P}$, Count of data per cluster $c$}
\leftline{\textbf{Output}: Updated subpocket prototypes $\mathcal{P}$}
\begin{algorithmic}[1]
   \STATE $S$ = FindNearest($\rm{H}, \mathcal{P}$)
   \STATE $\mathcal{P} \leftarrow c\cdot \mathcal{P}\cdot \gamma + S^\top \rm{H}\cdot(1-\gamma)$
   \STATE $c \leftarrow c\cdot \gamma + S^\top \bm{1}\cdot(1-\gamma)$
   \STATE $\mathcal{P} \leftarrow \mathcal{P}/c$
\end{algorithmic}
\end{algorithm}
\subsection{Prototype-augmented Motif Generation}
\label{generation}
The intuition of this Prototype-augmented Motif Generation is that according to the similarity principle: \textbf{molecular motifs originating from similar subpockets are likely to bind with the target protein pocket with high affinity}.

In the generation process, we firstly obtain atom and residue embeddings from the hierarchical context encoder. The subpocket embedding $\vh_c$ can be obtained by sum pooling all the residue embeddings within 6 {\AA} of the focal atom.  To leverage the knowledge from the global interaction graph, we take a global information fusion step in Figure. \ref{illustration}(b): we add edges between the subpocket embedding with $K_p$ most similar subpocket prototypes in the global interaction graph with edge weights set as 1. Then we use a weighted graph neural network to propagate the global information and take the output subpocket representation as $\hat \vh_c$. The related subpocket prototype-motif interaction information can be encoded in $\hat \vh_c$ for the next motif prediction.
We show the details of the weighted GNN in the Appendix. \ref{interaction graph}. 


\textbf{Focal Motif Prediction}: Before predicting the next motif, we first select a focal motif which the next motif attaches with. Two atom-wise MLPs are used as classifiers: protein atom classifier (for $t = 1$) and molecular atom classifier (for $t \ge 2$).
(1) At $t = 1$, all the known context information is the protein pocket. The protein atom classifier takes the hidden representations of protein atoms as input, and predicts whether new ligand atoms can be generated within 4 {\AA}. (2) For $t \geq 2$, the molecule atom classifier selects a focal atom from the ligand atoms generated in the previous $t-1$ steps. The motif that the focal atom belongs to is chosen as the focal motif. If no atom/motif is selected as focal, the generation process is completed. 

\textbf{Next Motif Prediction}: Given the focal motif $\mathcal{M}_f$, the label of the next motif is predicted as:
\begin{equation}
    P_m= \mathop{\rm softmax}\limits_{\mathcal{M}\in V_{\mathcal{M}}}({\rm MLP}^{\mathcal{M}}({\rm e}(\mathcal{M}_f), \sum_{i \in \mathcal{M}_f} \vh_i, \hat \vh_c)\cdot {\rm e}(\mathcal{M}))
    \label{next motif pred}
\end{equation}
where $P_m$ is the distribution over the motif vocabulary $V_\mathcal{M}$, ${\rm e}(\mathcal{M})$ denotes the motif embedding, $\sum_{i \in \mathcal{M}_f} \vh_i$ is the sum of the atom embeddings in the focal motif, and $\hat{\vh}_c$ is the enriched subpocket representation. We use a MLP to fuse the context information and use a dot product to score each motif. 
Since there is no focal motif at the first step ($t = 1$), we regard \textit{no motif} as a special motif type and also learn its embedding in training. 

\textbf{Motif Attachment Prediction:}
With the predicted motif, the next step is to attach the new motif to the generated molecule. Such a step is not deterministic since there are potentially several attachment configurations (See Figure.\ref{illustration}). Our goal here is to select the most appropriate attachment. Specifically, we enumerate different \textit{valid} attachments and form a candidate set $C$. 
We employ GIN \citep{xu2018powerful} to encode the candidate molecular graphs (${\rm GIN}(\cdot)$) and the probability $P_a$ of picking every molecule attachment is calculated as:
\begin{equation}
    P_a = \mathop{\rm softmax}\limits_{\mathcal{G}' \in C} ({\rm MLP}^{a} ({\rm GIN}(\mathcal{G}'), \hat \vh_{c})).
\end{equation}
We merge atoms or bonds in the process of motif attachment. By pruning chemically invalid molecules and merging isomorphic graphs with RDkit \cite{bento2020open}, we have $|C| \approx 3$ on the CrossDocked dataset. Therefore, the attachment prediction is also very efficient.

\textbf{Rotation Angle Prediciton}: 
As the flexibility of molecular structures largely lie in the degree of rotatable bonds \cite{jing2022torsional}, we focus on predicting the rotation angles in DrugGPS.
After attaching the new motif and obtaining the initial coordinates, we apply the encoder again to get the updated atom embeddings. Let $X,Y$ denote the two end atoms of the rotatable bond (let $Y$ denote the atom connecting the new motif).
We predict the change of the torsion angle $\Delta \alpha$:
\begin{equation}
    \Delta \alpha = {\rm MLP}^{\alpha}(\vh_X, \vh_Y, \vh_{\mathcal{G}}){\rm mod}2\pi,
\end{equation}
where $\vh_X$ and $\vh_Y$ indicate the embeddings of $X$ and $Y$; $\vh_{\mathcal{G}}$ denotes the embedding of the molecule, which is obtained with a sum pooling. $\Delta \alpha$ is also rotationally and translationally invariant since the prediction is based on the representations from the equivariant encoder.
Finally, the coordinates of the atoms in the new motif are updated by rotating $\Delta \alpha$ around line $XY$. As for the first motif in the generation, we use a distance-based initialization for its coordinates as there is no reference ligand atoms. More details of the generation process are included in Appendix. \ref{interaction graph}.

\subsection{Model Training}
\label{training}
In the training stage, 
the motifs of molecules are randomly masked
and DrugGPS is trained to recover the masked ones. Specifically, for each pocket-ligand pair, we sample a mask ratio
from the uniform distribution $U[0, 1]$ and mask the corresponding number of molecular motifs. The generation of motifs is in a breadth-first order where the root motif is set as the motif closest to the pocket. The atoms with valence bonds to the masked motifs are defined as focal atom candidates. If all molecular atoms are masked, the focal atoms are defined as protein
atoms that have masked ligand atoms within 4 {\AA}.

For the focal atom/motif prediction, we use a binary cross
entropy loss $\mathcal{L}_{focal}$ for the classification of focal atoms.
For the motif type and attachment prediction, we use cross entropy losses for the classification,
denoted as $\mathcal{L}_{motif}$ and $\mathcal{L}_{attach}$.  
As for the torsion angle prediction, we fit angles with von Mises distributions with $\mathcal{L}_\alpha$ following \citep{senior2020improved}. For the distance-based initialization, we minimize an MSE loss $\mathcal{L}_{d}$ with respect to the pairwise distances.
In the training process, we aim to minimize the sum of the above loss functions:
\begin{equation}
    \mathcal{L} = \mathcal{L}_{focal}+ \mathcal{L}_{motif} +\mathcal{L}_{attach} +\mathcal{L}_{\alpha} + \mathcal{L}_{d}.
\end{equation}

\section{Experiments}
\subsection{Experimental Settings}
\textbf{Dataset:}
Following previous works \citep{peng2022pocket2mol, liu2022generating},
we use the CrossDocked dataset \citep{francoeur2020three}
which contains 22.5 million protein-molecule pairs.
We filter out data points whose
binding pose RMSD is greater than 1 {\AA}, leading to a refined subset with around 180k data points. 
We consider two data splitting schemes to test the generalization abilities of models: (1) \textbf{Sequence-based Clustered Split (SCS)}  uses mmseqs2 \citep{steinegger2017mmseqs2} to cluster data at 30$\%$ sequence identity and (2) \textbf{Pocket-based Clustered Split (PCS)}  uses PocketMatch \cite{yeturu2008pocketmatch} to cluster data with a similarity threshold of 0.75. Specifically, PocketMatch represents pockets in a frame-invariant manner and compares pairs of sites based on the alignment of sorted distance sequences and the residue types. 
For both data splits, we randomly draw 100,000
protein-ligand pairs for training and 100 proteins from remaining clusters for testing. Therefore, the training and the testing set contain sequentially or structurally different pockets.
For evaluation, 100 molecules are randomly sampled for each protein pocket in the test set. More details of the dataset split are shown in Appendix. \ref{split details}.

\textbf{Baselines:} 
DrugGPS is compared with five state-of-the-art baseline methods including LiGAN \cite{ragoza2022generating}, AR \cite{luo20213d}, GraphBP \cite{liu2022generating}, Pocket2Mol \cite{peng2022pocket2mol}, and our previous work FLAG \cite{zhang2023molecule}.

\begin{table*}[t]
\renewcommand{\arraystretch}{1.2}
\renewcommand\tabcolsep{2.9pt}
\centering
\scriptsize
\caption{Comparing the generated molecules' properties by different methods under the \textbf{pocket-based clustered split}. We report the means and standard deviations. The properties of the test set are shown for reference and the best results are bolded.}
\vskip 0.15in
\begin{tabular}{c|ccccccccc}
\toprule
     \makecell[c]{Methods}& \makecell[c]{Vina Score\\(kcal/mol, ↓)} & \makecell[c]{High\\ Affinity(↑)} & \makecell[c]{QED (↑)} & SA (↑) & LogP & \makecell[c]{Lip. (↑)}  &\makecell[c]{Sim. Train (↓)} &\makecell[c]{Div. (↑)} &\makecell[c]{Time (↓)} \\ \midrule
     
    Testset&-7.145$\pm$2.24&- &0.465$\pm$0.25 &0.736$\pm$0.12&0.941$\pm$2.25&4.468$\pm$1.54&-&-&- \\
    
    LiGAN &\makecell[c]{-6.032$\pm$1.89} & \makecell[c]{0.194$\pm$0.26} & \makecell[c]{0.365$\pm$0.27}&\makecell[c]{0.615$\pm$0.20}&\makecell[c]{-0.015$\pm$2.48}
    & \makecell[c]{4.002$\pm$0.92} & \makecell[c]{0.410$\pm$0.22} & \makecell[c]{0.667$\pm$0.15}&\makecell[c]{1819.8$\pm$560.7} \\ 
    
    AR& \makecell[c]{-6.114$\pm$1.66} & \makecell[c]{0.235$\pm$0.23} & \makecell[c]{0.483$\pm$0.18} & \makecell[c]{0.662$\pm$0.19}&\makecell[c]{0.210$\pm$1.76}
    &\makecell[c]{4.688$\pm$0.45}&\makecell[c]{0.394$\pm$0.21}&\makecell[c]{0.650$\pm$0.13}&\makecell[c]{15986.4$\pm$9851.0}
    \\ 
    
    \makecell[c]{GraphBP} &\makecell[c]{-6.745$\pm$1.82} & \makecell[c]{0.378$\pm$0.29} & \makecell[c]{0.455$\pm$0.19} & \makecell[c]{0.710$\pm$0.18} & \makecell[c]{0.457$\pm$2.10}&
    \makecell[c]{4.783$\pm$0.34}&
    \makecell[c]{0.378$\pm$0.26}&
    \makecell[c]{0.659$\pm$0.12}&\makecell[c]{1162.8$\pm$438.5}\\ 
    
    \makecell[c]{Pocket2Mol} &  \makecell[c]{-6.869$\pm$2.19} & \makecell[c]{0.413$\pm$0.23} & \makecell[c]{0.524$\pm$0.24} & \makecell[c]{0.726$\pm$0.21} & \makecell[c]{0.830$\pm$2.17}&
    \makecell[c]{4.892$\pm$0.22}&
    \makecell[c]{0.364$\pm$0.19}&
    \makecell[c]{0.695$\pm$0.17}&\makecell[c]{2827.3$\pm$1456.8}\\

    \makecell[c]{FLAG} &  \makecell[c]{-6.956$\pm$1.92} & \makecell[c]{0.445$\pm$0.22} & \makecell[c]{0.552$\pm$0.20} & \makecell[c]{0.737$\pm$0.19} & \makecell[c]{0.745$\pm$2.09}&
    \makecell[c]{4.904$\pm$0.14}&
    \makecell[c]{0.388$\pm$0.18}&
    \makecell[c]{\textbf{0.704$\pm$0.18}}&\makecell[c]{1289.1$\pm$378.0}\\
    
    \makecell[c]{DrugGPS} & \makecell[c]{\textbf{-7.276$\pm$2.14}} & \makecell[c]{\textbf{0.565$\pm$0.23}} & \makecell[c]{\textbf{0.613$\pm$0.22}} & \makecell[c]{\textbf{0.743$\pm$0.18}}&\makecell[c]{0.913$\pm$2.15}&\makecell[c]{\textbf{4.917$\pm$0.12}}&\makecell[c]{\textbf{0.360$\pm$0.21}} &\makecell[c]{0.681$\pm$0.15}&\makecell[c]{\textbf{1007.8$\pm$554.1}} \\
\bottomrule
\end{tabular}
\vskip -0.1in
\label{main results}
\end{table*}

\textbf{Model:} The number of layers for the atom and residue-level encoder is set as 6 and 3 respectively. $K_a$ and $K_r$ are set as 32 and 8 respectively. The number of attention head $M$ is set as 4; The number of weighted GNN layers is 2.
The hidden dimension $d$ is set as 256. The threshold $\tau$ in motif extraction is set to 100 and $|V_\mathcal{M}| = 890$ in the default setting.
The number of subpocket prototypes is set as 128 in the default setting. We update the prototypes every 200 iterations with momentum $\gamma = 0.9$. 
In the global information fusion, the input subpocket embedding is linked with the top 4 closest prototypes measured by cosine similarity.  
The model is trained with the Adam optimizer with a learning rate of 0.0001. The batch size is 4 and the number of total training iterations is 1,000,000. 

\textbf{Metrics:}
We choose widely-used metrics in previous works \citep{peng2022pocket2mol, liu2022generating}
to evaluate the sampled molecules: 
(1) \textbf{Vina Score} calculates the binding affinity between the generated molecules and the protein pockets with QVina \citep{trott2010autodock, alhossary2015fast}; (2) \textbf{High Affinity} measures the percentage of pockets that have generated molecules with higher affinity than the references in the test set; (3) \textbf{QED} measures how likely a molecule is a potential drug candidate; (4) \textbf{Synthesizability (SA)} represents the difficulty of drug synthesis (normalized between 0 and 1 and higher values indicate easier synthesis); 
(5) \textbf{LogP} is the octanol-water partition coefficient and good drug candidates have LogP ranging from -0.4 to 5.6 \citep{ghose1999knowledge}; 
(6) \textbf{Lipinski (Lip.)} calculates how many rules the molecule obeys the Lipinski’s rule
of five \citep{lipinski2012experimental}; (7) \textbf{Sim. Train} represents the Tanimoto similarity \citep{bajusz2015tanimoto} with the most similar molecules in the training set; (8) \textbf{Diversity (Div.)} measures the diversity of generated molecules for a binding pocket. (9) \textbf{Time} records the time to generate 100 valid molecules for a pocket. 
All the generated molecules by different methods are optimized with universal force fields \citep{rappe1992uff}.

\begin{figure}[t]
	\centering
    \subfigure[]{\includegraphics[width=0.46\linewidth]{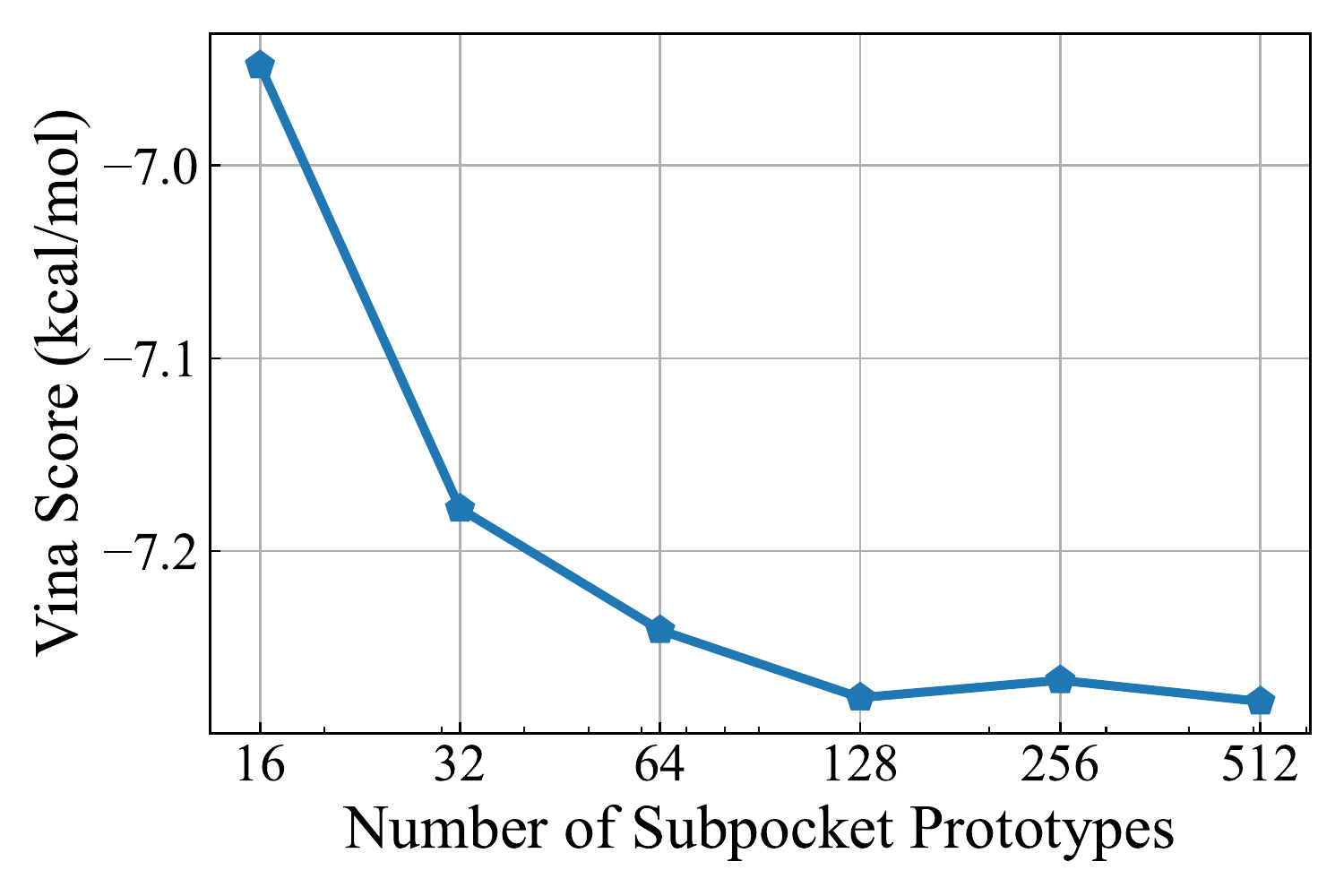}}
    \subfigure[]{\includegraphics[width=0.46\linewidth]{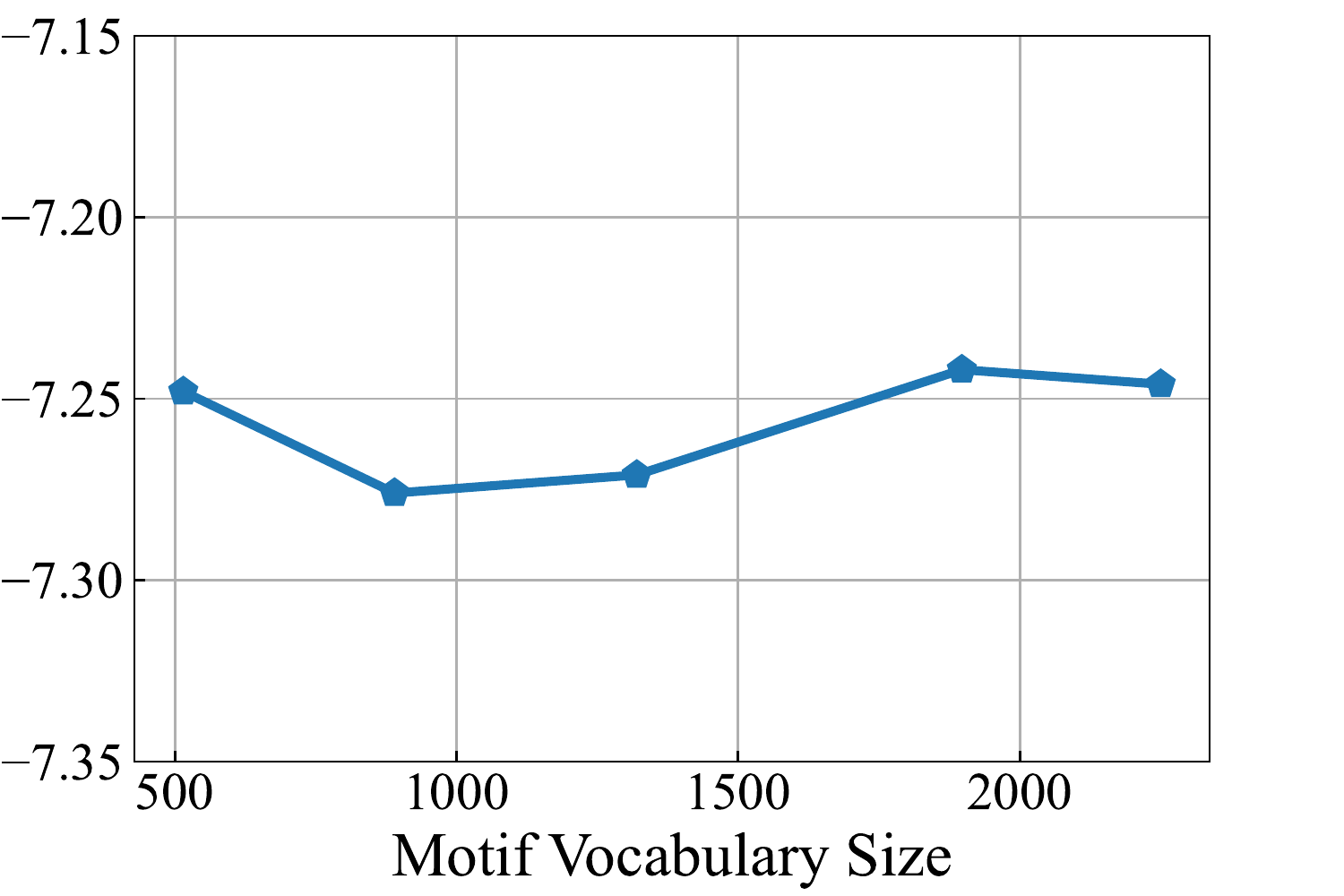}}
	\caption{Hyperparameter analysis with respect to the (a) number of subpocket prototypes $N$ and (b) motif vocabulary size $|V_\mathcal{M}|$.}
	\label{hyperparameter}
 \vspace{-1em}
\end{figure}
\subsection{Experimental Results}
We show the generated molecules' properties in Table. \ref{main results} (PCS) and Table. \ref{main results sequence based} (SCS) in Appendix. \ref{more results}. Generally, we find PCS more challenging:
the average vina scores of the generated molecules drop a lot for baseline methods from SCS to PCS (e.g., -7.288 to -6.869 for Pocket2Mol).
This is not surprising since proteins with overall low sequence similarities may still have similar pockets \cite{eguida2022estimating} in SCS, which helps generate high-affinity molecules for pockets in the test set.
Table. \ref{gap} also shows that PCS results in larger train-test performance gap.
We use pocket-based clustered split as the default setting in the rest of this paper to test the generalization abilities. 
Thanks to the constructed global interaction graph and prototype-augmented motif generation scheme, DrugGPS can still generate molecules with high affinity in PCS and does not drop much compared with SCS (-7.276 vs. -7.345).
Moreover, DrugGPS also manages to generate diverse molecules with high drug-likeliness and synthesizability, and with
low similarities to the molecules in the training dataset.
Note that 100$\%$ of the molecules are valid because DrugGPS explicitly filter out invalid candidates in the attachment selection step (A molecule is valid if it can be sanitized by RDkit \citep{bento2020open}). 
Finally, we compare the computational efficiency for molecule generation. With the motif-based generation scheme, DrugGPS can shorten the generation steps and is more efficient than baseline methods. More results and discussions are included in Appendix. \ref{more results}.

In Fig.\ref{ablation}(b), we also use t-SNE \cite{van2008visualizing} to visualize the sampled subpocket embeddings and their prototypes. 
We can observe that
the prototypes can mostly occupy the centers of subpocket embeddings,
which verifies the effectiveness of the learned subpocket prototype. 

\begin{figure}[t]
    \centering
    \subfigure[]{\includegraphics[width=0.46\linewidth]{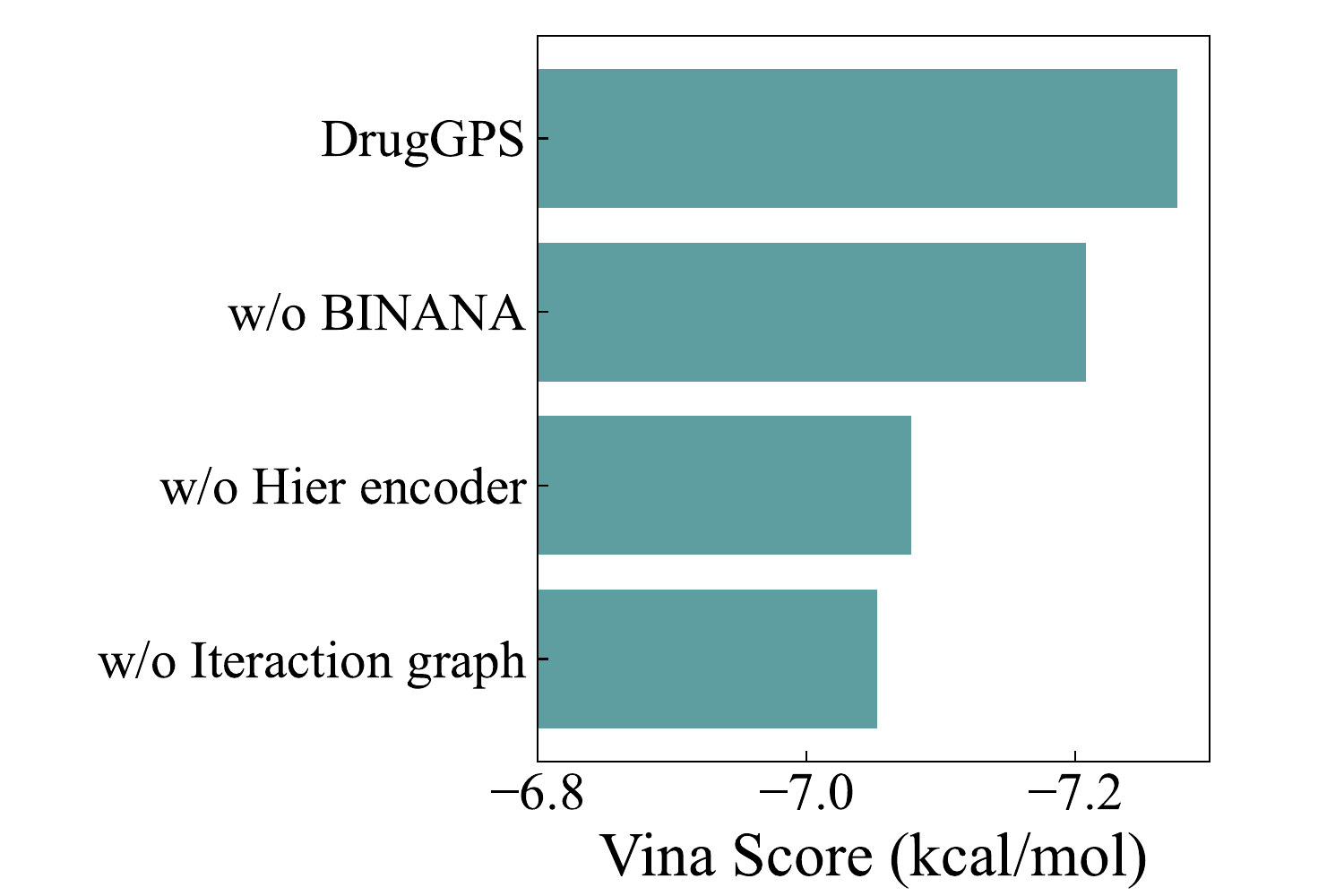}}
    \subfigure[]{\includegraphics[width=0.46\linewidth]{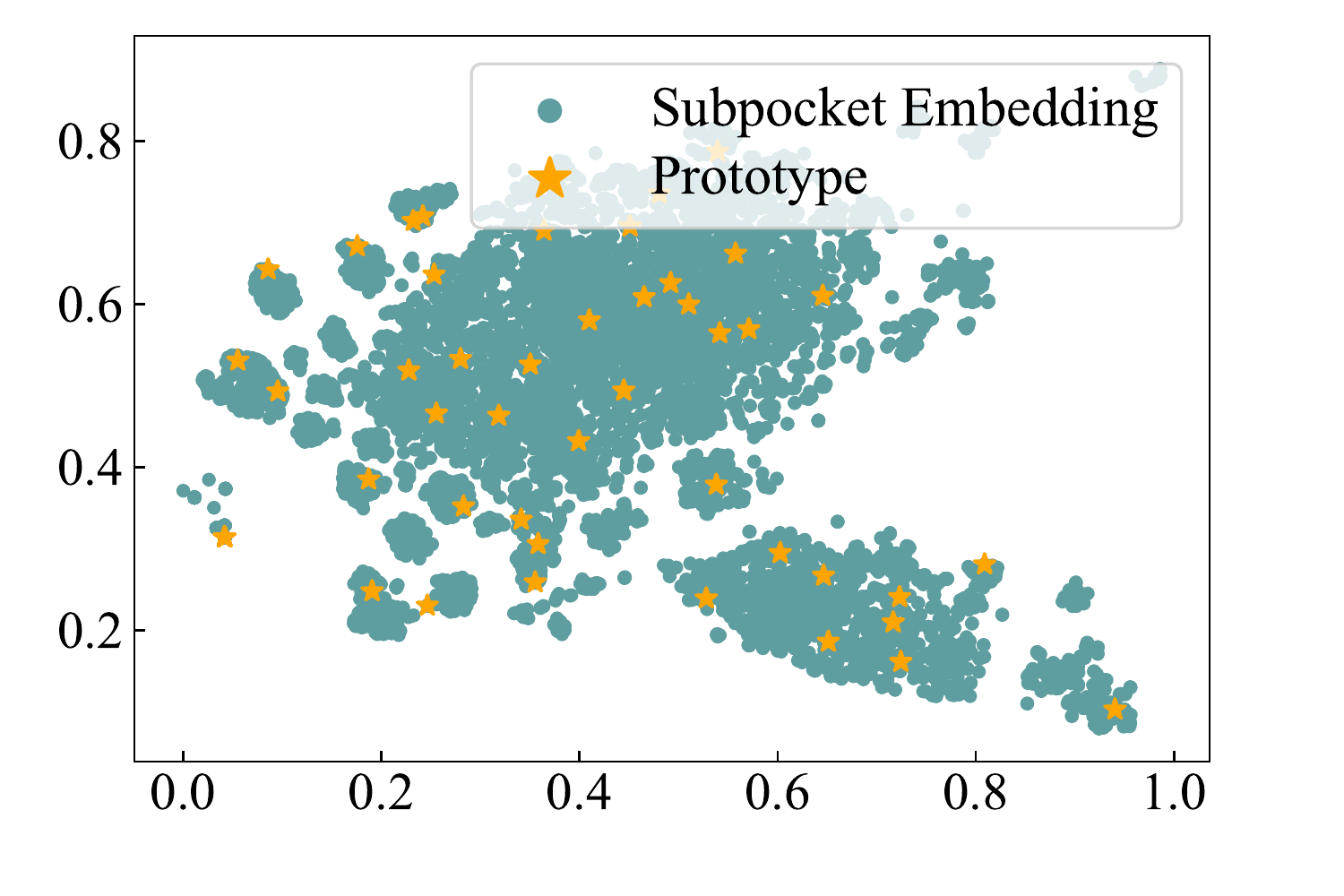}}
    \caption{(a) Ablation studies (b) t-SNE visualization of 10000 randomly sampled subpocket embeddings and 56 prototypes.}
    \vspace{-1em}
    \label{ablation}
\end{figure}

\begin{figure*}[t]
    \centering
    \includegraphics[width=0.98\linewidth]{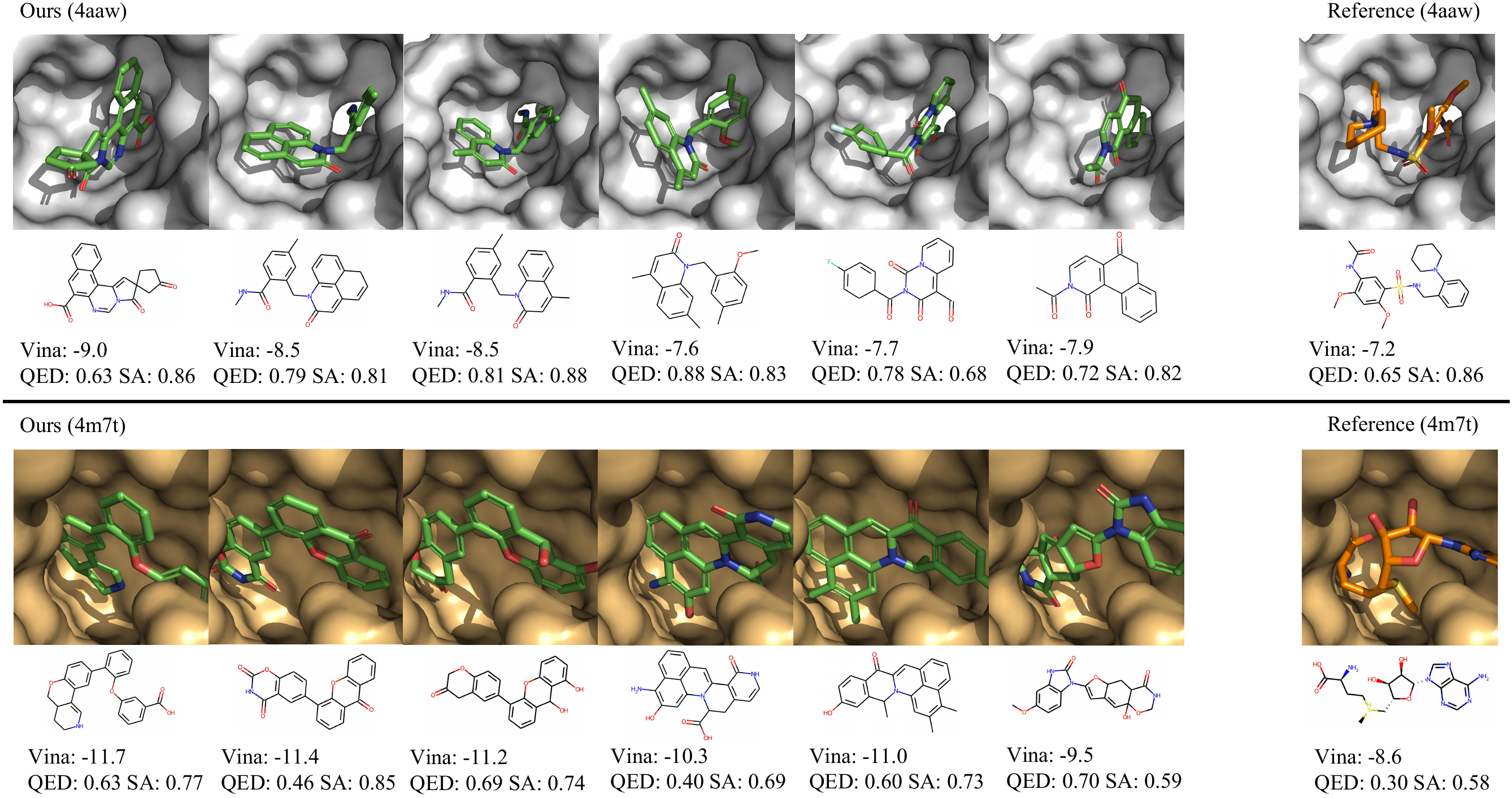}
    \caption{Examples of the generated molecules with higher binding affinities than the references. We report the vina scores, QED, and SA scores for each molecule. A lower Vina score indicates higher binding affinity.}
    \label{case}
\end{figure*}
\subsection{Case Studies}
In Figure \ref{case}, we further provide several generated molecule examples that have higher binding affinities (lower vina scores) than their corresponding reference molecules. 
Firstly, 
the generated molecules have novel structures that are different from the reference molecules. This implies that DrugGPS can generate novel and diverse structures.
Furthermore, the generated molecules also exhibit high QED and SA scores, showing their potential to be good drug candidates.
Finally, the generated molecules contain realistic substructures (e.g., benzene ring), which should attribute to the motif-based generation scheme. In DrugGPS, we focus on the prediction of rotatable bonds and use chemical tools to help determine the motif structures.  
We further provide quantitive substructure analysis in Appendix. \ref{more results}. 

\subsection{Hyperparameter Analysis $\&$ Ablation Studies}
In Fig. \ref{hyperparameter}, we explore the influence of two key hyperparameters: the number of subpocket prototypes $N$ and the motif vocabulary size $|V_\mathcal{M}|$. We can control hyperparameter $\tau$ to control $|V_\mathcal{M}|$. With the increase of $N$, the average vina scores of the generated molecules decrease (higher binding affinity) and gradually stabilize. This may be because a sufficient number of subpocket prototypes is required to represent the global distributions of subpocket embeddings. When it comes to the motif vocabulary size, we find an appropriate size of $|V_\mathcal{M}|$ is beneficial to generate high-quality ligand molecules. Too small motif vocabulary may limits generating large complex molecules while too large vocabulary may inhibit learning good motif representations. 

We further perform ablation studies to show the effectiveness of different modules in DrugGPS (Fig. \ref{ablation}(a)). Specifically, we remove the binding analysis tool BINANA, the residue-level encoder, and the global interaction graph in DrugGPS respectively as ``w/o BINANA'', ``w/o Hier encoder'', and ``w/o Interaction graph''. 
We find that removing these modules, especially the hierarchical encoder and global interaction graph, indeed degrades the performance of DrugGPS. 
This verifies that the necessity to capture the hierarchical context information and the effectiveness of the global interaction graph for generalizable ligand generation. 

\section{Conclusion}
In this paper, we propose DrugGPS, a generalizable structure-based drug design method. Inspired by the biochemical prior of subpockets, we propose a subpocket prototype-augmented ligand molecule generation scheme that levrages the global interaction knowledge of the whole dataset.
A hierarchical 3D graph transformer is also proposed to encode both the atom-level and residue-level information.
Experiments show that our model consistently outperforms baselines in generating realistic drug candidates with high affinities in challenging out-of-distribution settings.
Future works may include further leveraging interpretable and robust machine learning techniques \cite{zhang2023backdoor, zhang2022fldetector, zhang2022model, zhang2022protgnn}, unifying protein and molecule pre-training \cite{zhang2021motif}, and extending our framework to other domains such
as protein design \cite{gao2022pifold}.

\nocite{langley00}

\section{Acknowledgements}
This research was partially supported by a grant from the National Natural Science Foundation of China (Grant No. 61922073).

\bibliography{main}
\bibliographystyle{icml2023}

\newpage
\appendix
\onecolumn
\section{More Details of Hierarchical Graph Transformer}
\label{model architecture}
Our hierarchical encoder includes the atom-level encoder and the residue-level encoder. The powerful 3D Graph Transformer is used as the model backbone. Here we first show more details of the graph transformer architecture. Then wen show the residue-level node and edge features following previous works \cite{ingraham2019generative, jin2022antibody}.

\textbf{Graph transformer architecture.} The atom/residue-level encoder contains $L$ graph transformer layers. Let $\vh^{(l)}$ be the set of node representations at the $l$-th layer. In each graph transformer layer, there are a multi-head
self-attention (MHA) and a feed-forward block (FFN). The layer normalization (LN) is applied before the two blocks \cite{xiong2020layer}. The details of MHA has been shown in Sec.\ref{encoder} and the graph transformer layer is formally characterized as:
\begin{align}
        \vh'^{(l-1)} &= {\rm MHA(LN(}\vh^{(l-1)})) + \vh^{(l-1)} \\
        \vh^{(l)} &= {\rm FFN(LN(}\vh'^{(l-1)})) + \vh'^{(l-1)}, \;\; (0 \leq l < L).
\end{align}

\textbf{Residue-level node features.} 
The residue-level encoder only keeps the C$_\alpha$ atoms to represent residues and constructs a $K_r$ nearest neighbor graph at the residue level.
Each residue node is represented by six features: polarity $f_p\in\set{0,1}$, hydropathy $f_h\in [-4.5, 4.5]$, volume $f_v\in [60.1, 227.8]$, charge $f_c\in \set{-1,0,1}$, and whether it is a hydrogen bond donor $f_d\in\set{0,1}$ or acceptor $f_a\in\set{0,1}$. We expand hydropathy and volume features into radial basis with interval sizes 0.1 and 10, respectively. Overall, the dimension of the residue-level feature vector $\vf_i$ is 112.

\textbf{Residue-level edge features.}
For each $i$-th residue, we let $x_i$ denote the coordinate of its  C$_\alpha$ and define its local coordinate frame $\mO_i = [\vc_i, \vn_i, \vc_i \times \vn_i]$ as:
\begin{equation}
    \vu_i = \frac{\vx_i - \vx_{i-1}}{\lVert \vx_i - \vx_{i-1} \rVert}, \quad
    \vc_i = \frac{\vu_i - \vu_{i+1}}{\lVert \vu_i - \vu_{i+1} \rVert}, \quad
    \vn_i = \frac{\vu_i \times \vu_{i+1}}{\lVert \vu_i \times \vu_{i+1} \rVert}.
\end{equation}
Based on the local frame, the edge features between residues $i$ and $j$ can be computed as:
\begin{equation}
    \bm{e}_{ij}^{res} = {\rm Concat}\bigg(
    E_{\mathrm{pos}}(i - j), \quad
    \mathrm{RBF}(\Vert \vx_{i} - \vx_{j} \Vert),  \quad
    \mO_i^\top \frac{\vx_{j} - \vx_{i}}{\Vert \vx_{i} - \vx_{j} \Vert},\quad
    \vq(\mO_i^\top \mO_j)
    \bigg). \label{eq:edge-feature}
\end{equation}
The edge feature $\bm{e}_{ij}^{res}$ contains four parts. The positional encoding $E_{\mathrm{pos}}(i - j)$ encodes the relative sequence distance between two residues. 
The second term $\mathrm{RBF}(\cdot)$ is a distance encoding with radial basis functions. 
The third term is a direction encoding corresponding to the relative direction of $\vx_j$ in the local frame of $i$-th residue. 
The last term $\vq(\mO_i^\top \mO_j)$ is the orientation encoding of the quaternion representation $\vq(\cdot)$ of the spatial rotation matrix $\mO_i^\top \mO_j$ \cite{huynh2009metrics}. 
Overall, the dimension of the residue-level edge feature $\bm{e}_{ij}^{res}$ is 39.

\section{More Details of Prototype-augmented Motif Generation}
\label{interaction graph}
\textbf{Global interaction graph construction.} The number of subpocket prototypes is set as 128 in the default setting. 
For the stability of optimization, the encoder first goes through a warm-up period with 50,000 iterations. After the warm-up, the prototypes are first initialized with K-Means and further updated every 200 iterations with momentum $\gamma = 0.9$. The edge weights $W_{ij}$ are updated in the interval of 10,000 iterations. 
DrugGPS is validated every 10,000 iterations and we take the checkpoint with the lowest validation loss for ligand generation.
For the input subpocket, We
add edges between the subpocket embedding with 4 most
similar subpocket prototypes measured by cosine similarity.

\textbf{Global information fusion with Weighted GNN.}
To enrich the subpocket representation with the global interaction information, we use a weighted GNN to propagate information on the constructed interaction graph. The subpocket prototypes and motif embeddings are firstly converted to the same hidden space with learnable transformation matrices. We use a weighted GNN adapted from GIN \cite{xu2018powerful} for information propagation and aggregation:
\begin{equation}
    x_v^l= {\rm MLP}(x_v^{l-1} + \frac{1}{\sum_{u \in \mathcal{N}(v)} {W}_{uv}}\sum_{u \in \mathcal{N}(v)}{W}_{uv}x_u^{l-1}),
\end{equation}
where $x_v^l$ denotes node $v$'s embedding at the $l$-th layer and $\mathcal{N}(v)$ denotes its neighbors. ${W}_{uv}$ is the edge weight calculated in Equation. \ref{edge weight}. Finally, we take the input subpocket embedding at the final layer as $\hat \vh_c$. In prototype-augmented motif generation, $\hat \vh_c$ is leveraged in Equation. \ref{next motif pred} for next motif prediction. It not only encodes the residue-level context information, but capture the similar subpocket prototype-molecular motif interaction as well.
Therefore, the next motif prediction classifier can learn to give motifs with high binding affinity higher scores.

\textbf{Distance-based initialization.}
To decide the 3D positions of the generate motifs, it is challenging for the first motif as there is no reference ligand atoms.
Following \citep{jin2022antibody}, we use a distance-based initialization strategy to determine the position of the first motif, which is more accurate and stable than random initialization. Specifically, a distance matrix $\mD \in \mathbb{R}^{(n'+m')\times(n'+m')}$ is set as:
\begin{equation}
    \mD_{i,j} =
\begin{cases}
    \Vert \bm{r}_i - \bm{r}_j \Vert & i,j \leq n' \\
    {\rm MLP}^d(\bm{h}^{(0)}_i, \bm{h}^{(0)}_j) & i \leq n', j > n' \\
    \Vert \bm{r}_i - \bm{r}_j \Vert & i, j > n',\\
\end{cases}
\end{equation}
where $n'$ and $m'$ denote the number of sampled protein atoms for reference and the number of atoms in the first molecular motif. $\bm{r}_i$ is the 3D coordinate of the atom. The distances within the protein atoms and motif atoms can be directly calculated. For the distances between molecular and protein atoms, we use ${\rm MLP}^d$ for prediction with the pairwise atom attributes as the input.
With the distance matrix $\mD$, we can obtain the coordinates of atoms by eigenvalue decomposition of its Gram matrix \citep{crippen1978stable}:
\begin{equation}
    \tilde{\mD}_{i,j} = 0.5(\mD_{i,1}^2 + \mD_{1,j}^2 - \mD_{i,j}^2), \quad \tilde{\mD} = \mU\mS\mU^\top
\end{equation}
where $\mS$ is a diagonal matrix with eigenvalues in descending order. The coordinate of each atom $\bm{r}_i$ is calculated as:
\begin{equation}
    \tilde{\vr}_i = [\mX_{i,1}, \mX_{i,2}, \mX_{i,3}], \quad \mX = \mU\sqrt{\mS}.
\end{equation}
Note that even though the predicted coordinates $\{\tilde{\vr}_i\}$ retain the original distance $\mD$, they are located in a different coordinate system. Therefore, we apply the Kabsch algorithm \citep{kabsch1976solution} to find a rigid body transformation $\{\mR, \vt\}$ that aligns the predicted protein coordinates $\{\tilde{\vr}_{1}, \cdots, \tilde{\vr}_{n'}\}$ with the reference coordinates $\{\vr_1, \cdots, \vr_{n'}\}$. Lastly, the coordinates of the first motif are calculated as:
\begin{equation}
    \vr_i = \mR \tilde{\vr}_{i} + \vt, ~i > n'.
\end{equation}
For generation steps with $t>1$, the coordinates of the attached motifs are determined and aligned similarly with RDkit \citep{bento2020open} and Kabsch algorithm \citep{kabsch1976solution}.

\textbf{Rotation matrix.}
In the generation of new motifs, if the focal motif is rotatable and the rotation angle $\Delta \alpha$ is known, we use the following rotation matrix $R_{3\times 3}$ and the translation vector $t_{3\times 1}$ to update the coordinates of the new motif. Let $X, Y$ denote the two end atoms of the rotatable bond ($Y$ denote the atom connecting the new motif) and ${\bm r}_X$ and ${\bm r}_Y$ be their coordinates. Let ${\bm n}$ denotes the normalized directional vector $\frac{{\bm r}_Y-{\bm r}_X}{\|{\bm r}_Y-{\bm r}_X\|}$ and $n_x, n_y$ and $n_z$ be its components along the $x, y$ and $z$ axis. Let $x_0, y_0$, and $z_0$ be the three components of ${\bm r}_X$. The rotation matrix and translation vector are:
\begin{equation}
    R_{3\times3}=\left[
  \begin{array}{ccc}   
    n_x^2K+cos(\Delta \alpha) & n_x n_yK - n_z sin(\Delta \alpha) & n_x n_z K +n_y sin(\Delta \alpha)\\  
    n_xn_yK+n_z sin(\Delta \alpha) & n_y^2 K + cos(\Delta\alpha) & n_yn_zK -n_x sin(\Delta \alpha)\\  
    n_xn_zK-n_ysin(\Delta \alpha)&n_yn_xK+n_x sin(\Delta \alpha)&n_z^2K + cos(\Delta \alpha)
  \end{array}
\right],
\end{equation}
\begin{equation}
    t_{3\times 1} = \left[
  \begin{array}{c}   
(x_0-n_x M)K+(n_z y_0 - n_y z_0)sin(\Delta \alpha)\\ 
(y_0-n_y M)K+(n_x z_0 - n_z x_0)sin(\Delta \alpha)\\ (z_0-n_z M)K+(n_y x_0 - n_x y_0)sin(\Delta \alpha)\\
  \end{array}
\right].
\end{equation}

Here, $K = 1- cos(\Delta \alpha)$ and $M = n_x x_0 + n_y y_0 + n_z z_0$. The coordinates ${\bm r}_i$ in the motif are updated as:
\begin{equation}
    {\bm r}_i' = R {\bm r}_i+t
\end{equation}

\section{More Details of Experimental Settings}
\textbf{More training and sampling details.} 
All the experiments are conducted on a NVIDIA Tesla
V100 GPU with 32G memory. It takes around 48 hours to train DrugGPS.
The implementation of DrugGPS is based on our previous work FLAG\footnote{\url{https://github.com/zaixizhang/FLAG}}.
To implement the baselines including LiGAN \footnote{\url{https://github.com/mattragoza/LiGAN}}, AR\footnote{\url{https://github.com/luost26/3D-Generative-SBDD}}, GraphBP\footnote{\url{https://github.com/divelab/GraphBP}}, and Pocket2Mol\footnote{\url{https://github.com/pengxingang/Pocket2Mol}}, we use the open-source codes following their default settings. DrugGPS and baseline methods are trained on the same data split for fair comparisons.

\textbf{More details of dataset split}
\label{split details}
We use two OOD data split in our work. 
(1) \textbf{Sequence-based Clustered Split} uses mmseqs2 \citep{steinegger2017mmseqs2} to cluster data at 30$\%$ sequence identity, which is a popular spliting scheme in previous works \cite{peng2022pocket2mol, luo2021autoregressive}. However, such splitting scheme only considers sequence similarity while neglecting
structural similarities, which is more important in structure-based drug design.
(2) \textbf{Pocket-based Clustered Split} uses PocketMatch \cite{yeturu2008pocketmatch} to cluster data with a similarity threshold of 0.75. The range of similarity score in PocketMatch is from 0 to 1 and higher scores indicate higher pocket similarities. 
PocketMatch compares binding pockets in a frame-invariant manner: each binding site is represented by 90 lists of sorted distances capturing the geometric and chemical properties of the pocket. The pocket pairs are then aligned based on distances and residue types to obtain similarity scores \cite{yeturu2008pocketmatch}.
Specifically in pocket-based clustered split, a set of initial centroids are sampled with pairwise similarity scores less than 0.75. The remaining pockets are compared with these centroids and assigned to the group with the higheset similarity score ($>$ 0.75). If there is no centroid that the pocket has an over 0.75 similarity score to, the pocket is selected as a new centroid. 

The output of the clustering algorithms are a set of clusters such that: (a) all centroids of clusters have similarity $< T$ to each other, and
(b) all members in a cluster have similarity $\ge T$ to the centroid. $T$ is the predefined similarity threshold. For both data splits, we randomly draw 100,000
protein-ligand pairs for training. The validation and test dataset are drawn from remaining clusters. Therefore, 
the training and test set contains sequentially or structurally different protein pockets. 


\section{More Experimental Results}
\label{more results}
\textbf{Ligand molecule generation under sequence-based clustered split. }We include the results under sequence-based clustered split in Table. \ref{main results sequence based}. This is the same split used in previous works \cite{peng2022pocket2mol, luo2021autoregressive} and we borrow part of the baseline results from \cite{peng2022pocket2mol}. We can observe that DrugGPS can also overperform baseline models on generating drug-like molecules with higher affinity in the popular SCS setting.
\begin{table*}[t]
\renewcommand{\arraystretch}{1.2}
\renewcommand\tabcolsep{2.9pt}
\centering
\scriptsize
\caption{Properties of the test set molecules and the generated molecules by different methods under the \textbf{sequence-based clustered split}. We report the means and standard deviations. This is the same split used in previous works \cite{peng2022pocket2mol, luo2021autoregressive} and we borrow part of the baseline results from \cite{peng2022pocket2mol}. The best results are bolded.}
\vskip 0.15in
\begin{tabular}{c|ccccccccc}
\toprule
     \makecell[c]{Methods}& \makecell[c]{Vina Score\\(kcal/mol, ↓)} & \makecell[c]{High\\ Affinity(↑)} & \makecell[c]{QED (↑)} & SA (↑) & LogP & \makecell[c]{Lip. (↑)}  &\makecell[c]{Sim. Train (↓)} &\makecell[c]{Div. (↑)} &\makecell[c]{Time (↓)} \\ \midrule
     
    Testset&-7.158$\pm$2.10&- &0.484$\pm$0.21 &0.732$\pm$0.14&0.947$\pm$2.65&4.367$\pm$1.14&-&-&- \\
    
    LiGAN &\makecell[c]{-6.114$\pm$1.57} & \makecell[c]{0.238$\pm$0.28} & \makecell[c]{0.369$\pm$0.22}&\makecell[c]{0.590$\pm$0.15}&\makecell[c]{-0.140$\pm$2.73}
    & \makecell[c]{4.027$\pm$1.38} & \makecell[c]{0.460$\pm$0.18} & \makecell[c]{0.654$\pm$0.12}&- \\ 
    
    AR& \makecell[c]{-6.215$\pm$1.54} & \makecell[c]{0.267$\pm$0.31} & \makecell[c]{0.502$\pm$0.17} & \makecell[c]{0.675$\pm$0.14}&\makecell[c]{0.257$\pm$2.01}
    &\makecell[c]{4.787$\pm$0.50}&\makecell[c]{0.409$\pm$0.19}&\makecell[c]{0.742$\pm$0.09}&\makecell[c]{19658.56$\pm$14704}
    \\ 
    
    \makecell[c]{GraphBP} &\makecell[c]{-7.132$\pm$1.75} & \makecell[c]{0.477$\pm$0.26} & \makecell[c]{0.516$\pm$0.15} & \makecell[c]{0.718$\pm$0.18} & \makecell[c]{0.442$\pm$2.08}&
    \makecell[c]{4.620$\pm$0.37}&
    \makecell[c]{0.415$\pm$0.24}&
    \makecell[c]{0.649$\pm$0.12}&\makecell[c]{1238.7$\pm$493.0}\\ 
    
    \makecell[c]{Pocket2Mol} &  \makecell[c]{-7.288$\pm$2.53} & \makecell[c]{0.542$\pm$0.32} & \makecell[c]{0.563$\pm$0.16} & \makecell[c]{\textbf{0.765$\pm$0.13}} & \makecell[c]{1.586$\pm$1.82}&
    \makecell[c]{4.902$\pm$0.42}&
    \makecell[c]{0.376$\pm$0.22}&
    \makecell[c]{0.688$\pm$0.14}&\makecell[c]{2503.51$\pm$2207}\\
    
    \makecell[c]{DrugGPS} & \makecell[c]{\textbf{-7.345$\pm$2.42}} & \makecell[c]{\textbf{0.620$\pm$0.29}} & \makecell[c]{\textbf{0.592$\pm$0.21}} & \makecell[c]{0.728$\pm$0.23}&\makecell[c]{1.134$\pm$2.26}&\makecell[c]{\textbf{4.923$\pm$0.11}}&\makecell[c]{\textbf{0.370$\pm$0.26}} &\makecell[c]{\textbf{0.695$\pm$0.17}}&\makecell[c]{\textbf{956.3$\pm$451.6}} \\
\bottomrule
\end{tabular}
\vskip -0.1in
\label{main results sequence based}
\end{table*}

\begin{figure*}[h]
    \centering
    \subfigure[Pocket-based clustered split]{\includegraphics[width=0.45\linewidth]{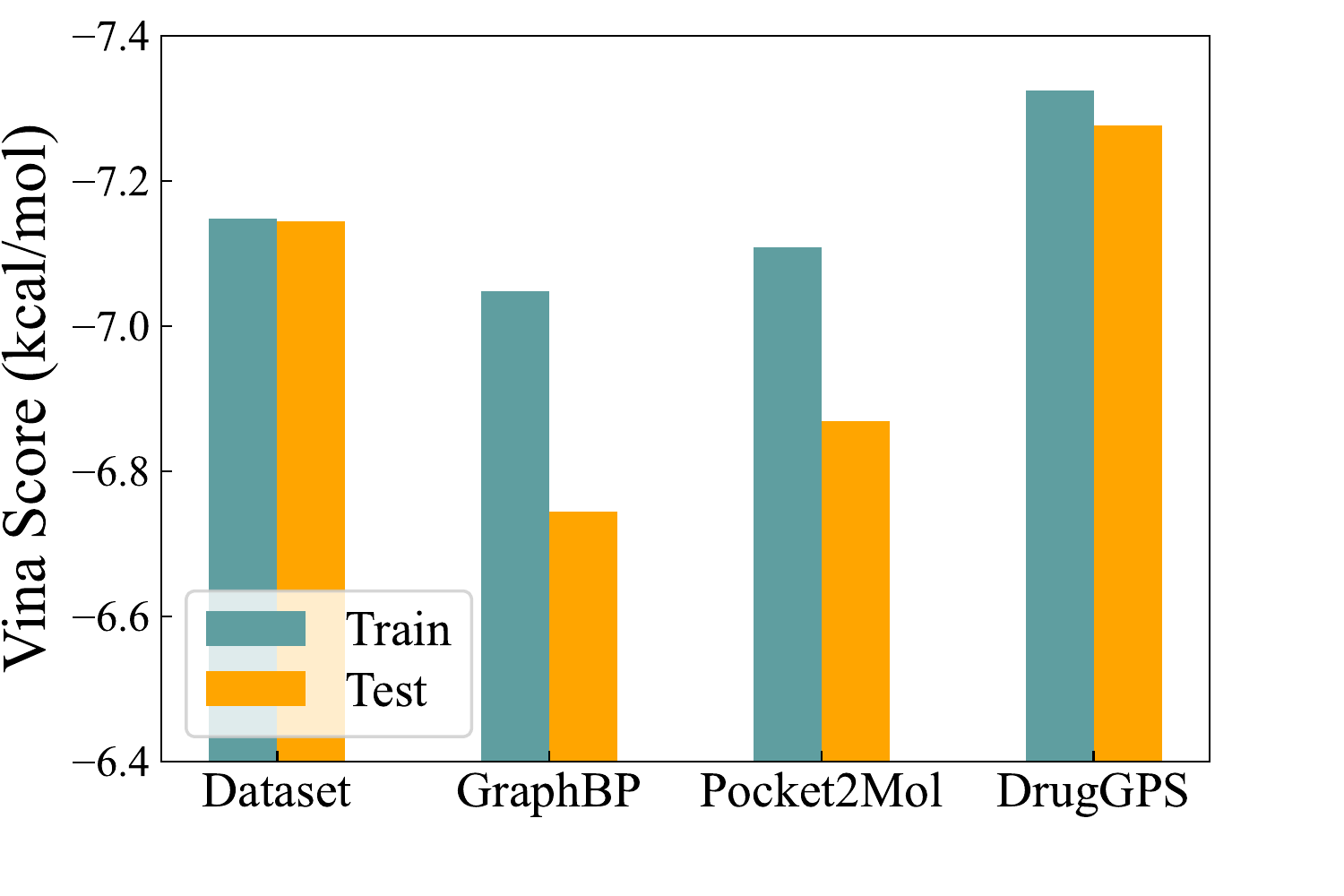}}
    \subfigure[Sequence-based clustered split]{\includegraphics[width=0.45\linewidth]{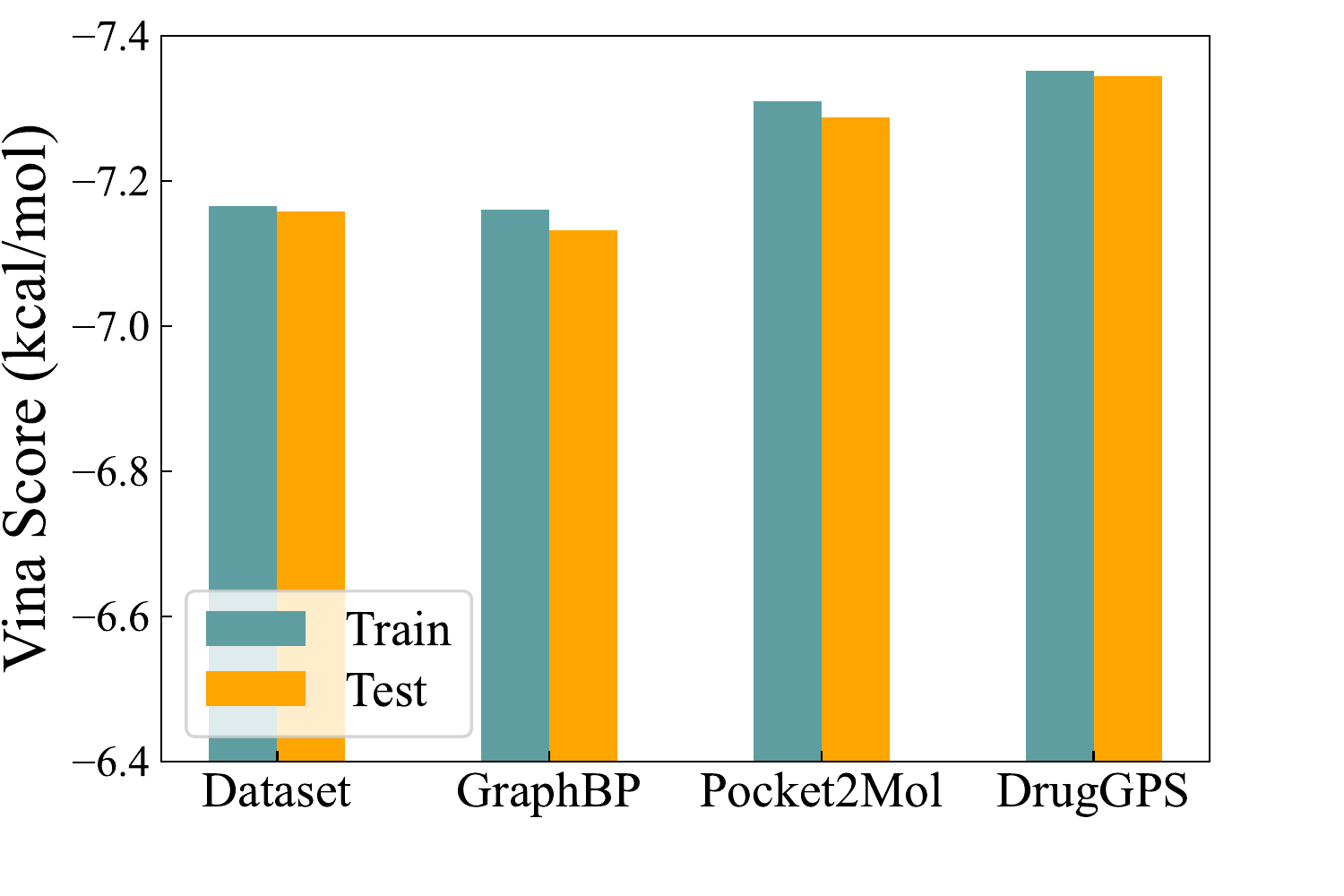}}
    \caption{Comparaing the Vina scores of the generated molecules on the training and testing dataset. (a) shows the pocket-based clustered split and (b) shows the sequence-based clustered split. A lower Vina score indicates higher binding affinity.}
    \label{gap}
\end{figure*}

\textbf{Peformance gap between training and testing set. }
In Table. \ref{gap}, we compare the Vina scores of the generated molecules on the training and testing dataset. For the training set, we randomly sample 100 proteins and generate 100 molecules for each target protein pocket similar to that of the test set. We compare the performance of our DrugGPS with two competitive baseline methods.
We also show the average vina scores from the sampled training set and the test set for reference.
The vina scores of the sampled training set and the test set are roughly the same.

Generally, we can observe that the pocket-based clustered split results in larger vina score gap and is more challenging. Compared with baselines, our DrugGPS has smaller gap, which indicates its better generalization ability.

\textbf{Qualitative substructure analysis. }We further show qualitative substructure analysis by calculating the KL divergence of the bond angles and dihedral angles between the molecules in the test set and the generated molecules in Table. \ref{kl}. We observe that DrugGPS can generate more realistic substructures than baseline methods (smaller angle distribution discrepancies between the generated molecules and the test set) due to its motif-based generation scheme. Moreover, we also compare DrugGPS with its variants and find the designed modules are also indispensable. Especially, the hierarchical encoder is quite important for DrugGPS to encode geometric information and predict rotation angles. 
\begin{table}
\caption{The KL divergence of the bond angles (upper part) and dihedral angles (lower part) between the molecules in the test set and the generated molecules by different methods. The lower letters represent the atoms in the aromatic rings. Pocket-based clustered split is used and the best results are bolded.}
\vskip 0.15in
\label{kl}
\centering
\footnotesize
\begin{tabular}{c|ccccccccc}
\toprule
    Angles & LiGAN & AR & GraphBP& \makecell[c]{Pocket2\\Mol} & \makecell[c]{FLAG}& \makecell[c]{w/o\\BINANA} & \makecell[c]{w/o Hier\\Encoder} & \makecell[c]{w/o Interaction\\Graph} & DrugGPS  \\ \midrule
    CCC & 7.16  & 2.31  &2.09& 0.93 &0.59&0.71 &0.86 &0.75 &\textbf{0.57}\\
    CCO & 7.68  & 2.19  &1.98& 0.94 &0.65&\textbf{0.62} &0.67 &0.64 &0.63  \\ 
    CCN  & 7.71  & 2.58  &2.46& 0.65 &\textbf{0.24}&0.27 &0.33 &0.26 &0.25 \\ 
    CNC & 6.39  & 3.28  &1.77& 0.72 &0.54&0.52 &0.79 &0.53 &\textbf{0.48}  \\ 
    OPO & 6.08  & 3.73  &3.71& \textbf{0.43} &0.57&0.48 &0.74 &0.49 &0.46 \\ 
    CC=O & 6.69  & 3.42  &3.47& 0.70 &0.48&0.47 &0.49 &0.48 &\textbf{0.45}  \\  \midrule
    cccc & 5.26 & 4.46 &3.62& 4.54 &\textbf{0.45}&0.46 &0.46 &\textbf{0.45} &\textbf{0.45}\\
    Cccc & 3.46 & 5.05 &2.28& 2.53 &1.41&1.40 &1.71 &1.34 &\textbf{1.32}\\
    CCCC & 4.14 & 2.15 &1.40& \textbf{0.69} &0.97&0.94 &1.53 &0.97 &0.92\\
    CCCO & 3.50 & 2.20 &1.25& 1.14 &0.96&0.91 &1.32 &0.88 &\textbf{0.87} \\
    OCCO & 2.14 & 2.16 &1.63&  1.73 &1.72&1.60 &1.88 &\textbf{1.52} &1.55\\
    CC=CC & 6.67 & 6.38 &3.44& 3.40 &2.20&2.15 &2.34 &2.08 &\textbf{2.06}\\
    \bottomrule
\end{tabular}
\end{table}

\begin{figure*}[t]
\centering
    \subfigure[]{\includegraphics[width=0.32\linewidth]{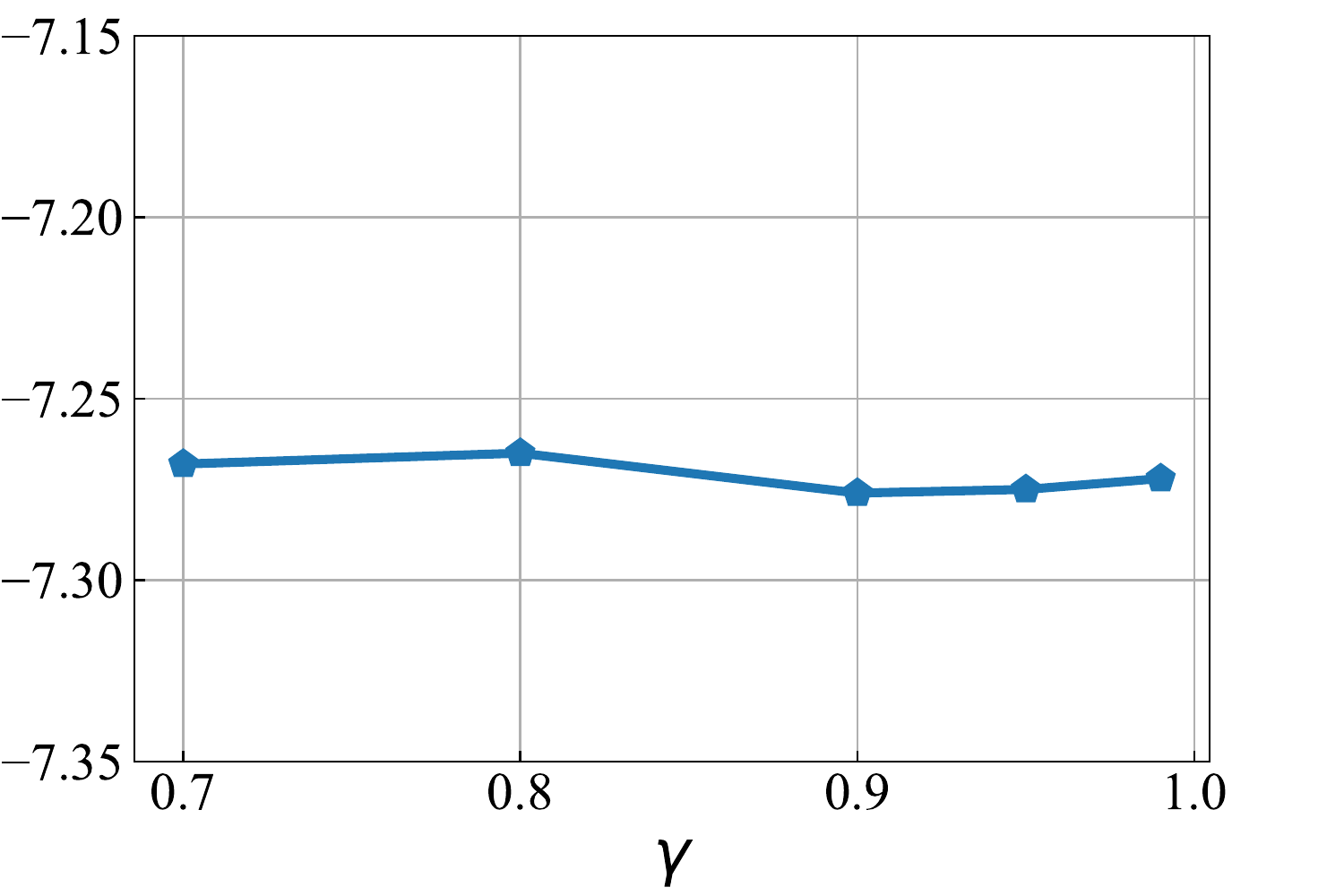}}
    \subfigure[]{\includegraphics[width=0.32\linewidth]{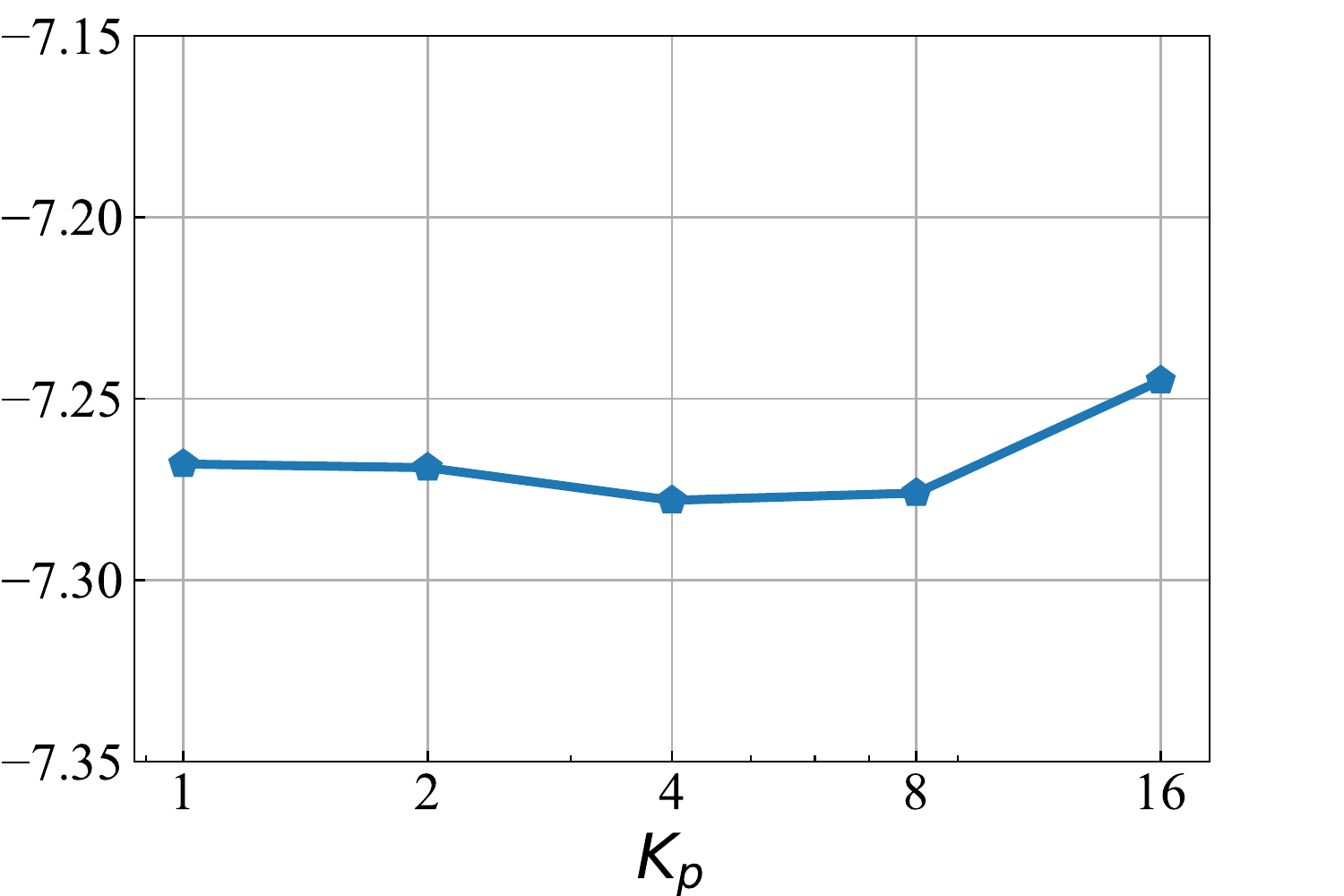}}
    \subfigure[]{\includegraphics[width=0.32\linewidth]{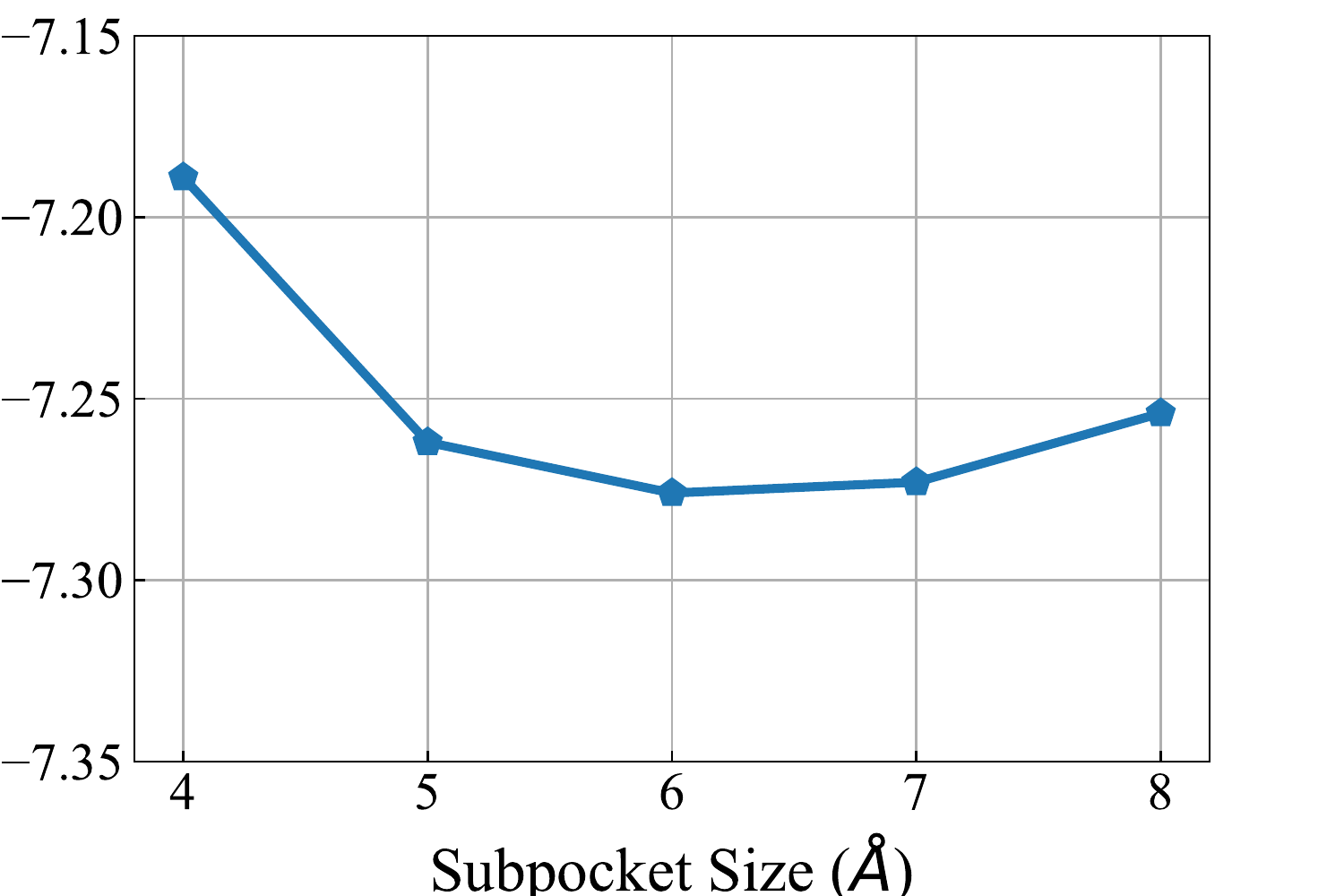}}
	\caption{Hyperparameter analysis with respect to (a) $\gamma$, (b) $K_p$: subpocket prototypes to link with in the global knowledge fusion, and (c) the subpocket size. DrugGPS is generally robust to the choice of hyperparameters. }
	\label{more hyperparameter}
\end{figure*}

\textbf{More Hyperparameter analysis. }We show more hyperparameter analysis with respect to $\gamma$, $K_p$ (number of subpocket prototypes to link with), and the subpocket size in Table. \ref{more hyperparameter}. We observe that DrugGPS is generally robust the choice of $\gamma$. The quality of generated molecules deteriorates slightly when too many subpocket prototypes are linked with in the global information fusion, which may lead to too much redundant information. Finally, we find choosing an appropriate subpocket size is important: some key residues may be neglected if the size is too small while some unrelated residue may be considered when the size is too large. In DrugGPS, we choose to sum up all the residue embeddings within 6 {\AA} to obtain subpocket representations in the default setting for the best performance. 

\begin{figure*}[t]
\centering
\includegraphics[width=0.98\linewidth]{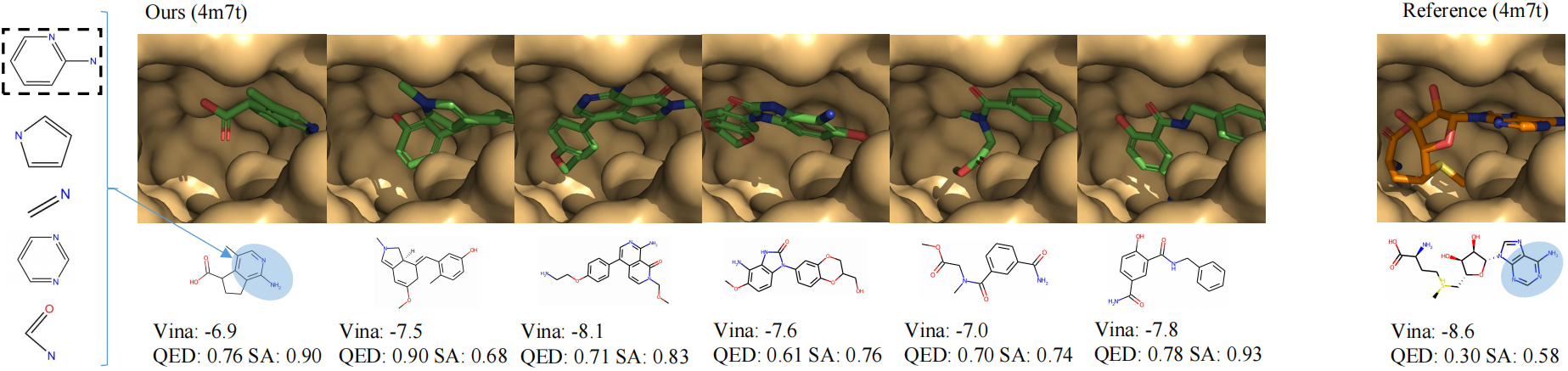}
	\caption{Generated examples with lower affinity than the reference ligands.}
	\label{failure}
\end{figure*}

\textbf{Failure Examples.} Here we show some examples that the generated molecules have lower affinity (higher Vina scores) than the reference in Figure. \ref{failure}. Therefore, these molecules have higher probabilities of not binding to the target pocket. The failure may be due to the following reasons: (1) Some generated molecules accidentally collide with the pocket, which is unrealistic in nature. (2) The auto-regressive generation scheme may limit its ability for overall optimization. We can observe that some generated molecules only occupy part of the pocket. In the future, we will tackle the aforementioned problems by designing penalty loss to prohibit colliding and explore generation schemes that facilitate overall optimization.

Structure-based drug design is a complicated conditional generation problem influenced by various factors (e.g., geometric and chemical constraints). It is common to have failure cases. However, even in the aforementioned failure situations, DrugGPS can still generate fragments aligning well with the global interaction information or the reference molecule. For example, for the first molecule in Figure 1, the first fragment generated by DrugGPS is quite similar to the reference molecule (marked with blue ovals). The fragment is also in the top-ranked fragments (listed on the leftmost) interacting with the corresponding subpocket prototype in the global interaction graph. Therefore, the prior that we impose into DrugGPS is generally helpful. 

\begin{figure*}[t]
\centering
\includegraphics[width=0.9\linewidth]{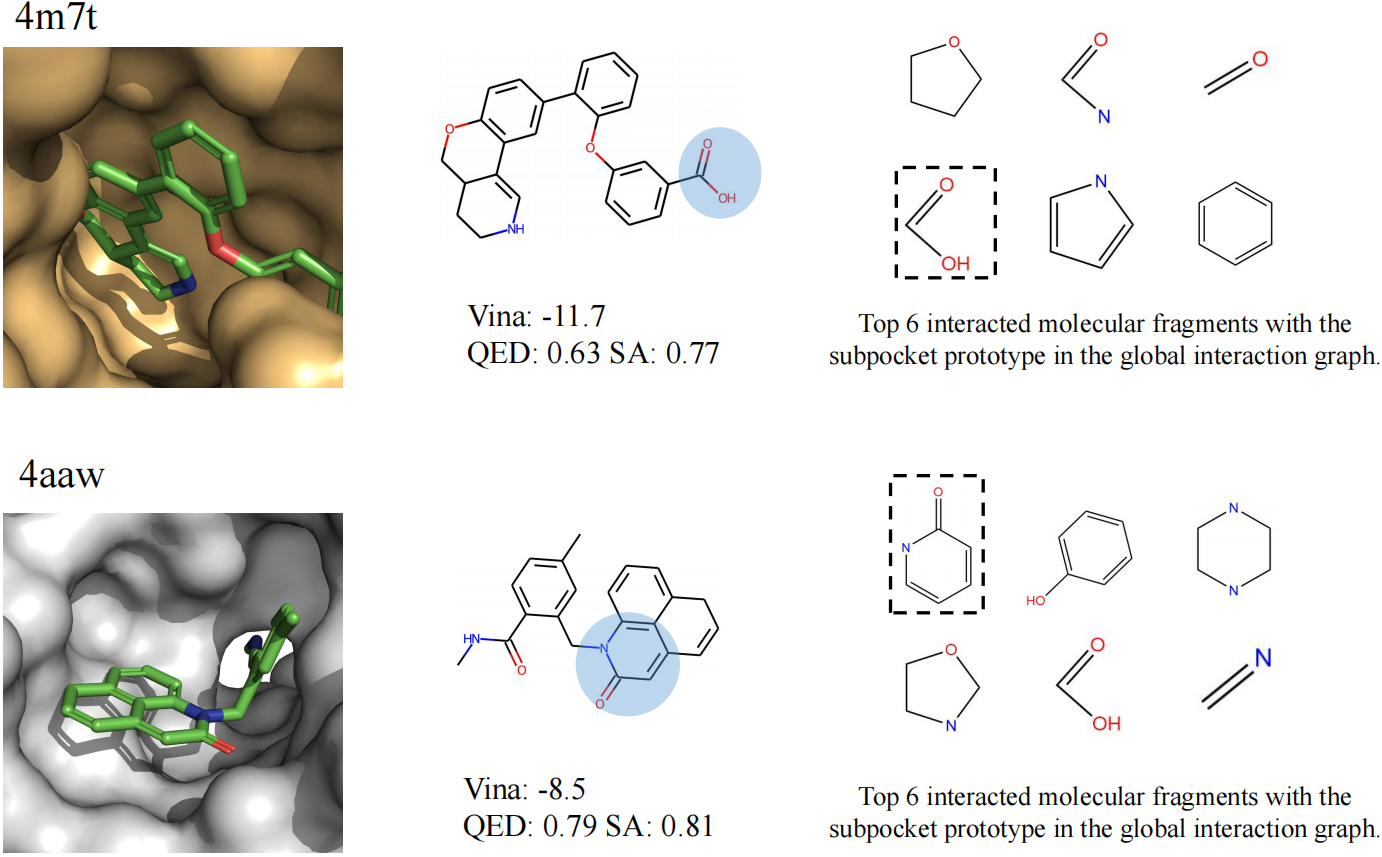}
	\caption{Global interaction helps ligand generation.}
	\label{motif_case}
\end{figure*}

\textbf{Global interaction helps ligand generation.}
To show the global interaction graph helps ligand generation, we include case studies showing the alignment between the generated ligand fragments and the fragment information derived from the global interaction graph in Figure. \ref{motif_case}. Specifically, the encoded subpocked representation is fed into the global interaction graph and further mapped to the closest subpocket prototype. The molecular motifs with the largest edge weights (Equ. \ref{edge weight}) with the subpocket prototype are shown here. In these cases, we observe that the fragment in the generated ligands (marked with blue ovals) are also within the top 6 interacted fragments from the global interaction graph (marked with black dashed boxes). These case studies support our claim that the global interaction graph helps fragment prediction and ligand generation.

\begin{figure*}[t]
\centering
\includegraphics[width=0.98\linewidth]{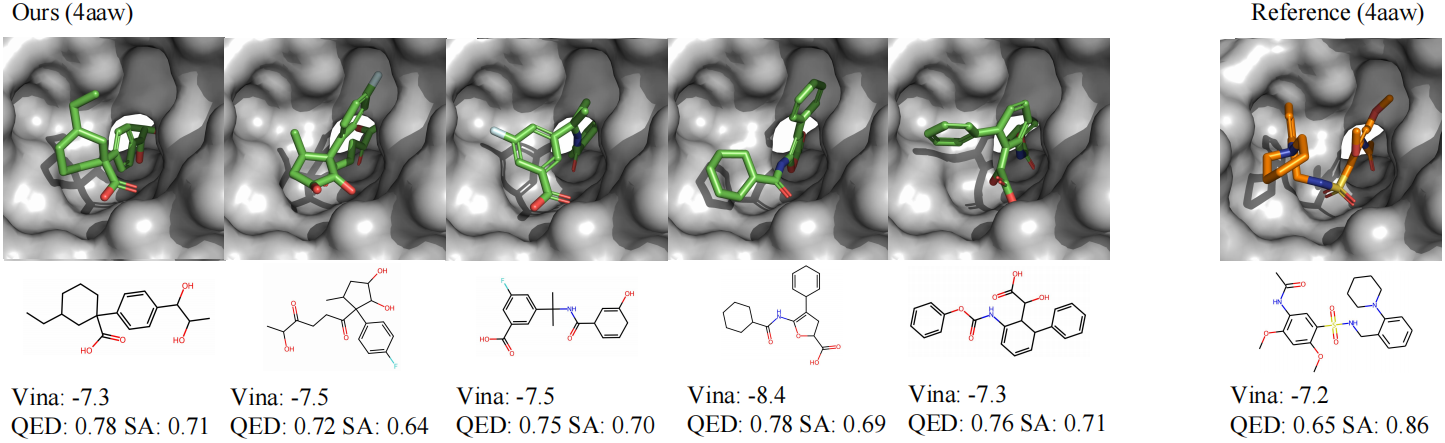}
	\caption{Generated molecules with monocyclic structures by the modified DrugGPS.}
	\label{monocyclic}
\end{figure*}

\textbf{The ratio of generated molecules with polycyclic structures.} In the CrossDocked dataset, 55.42$\%$ of ligand molecules have polycyclic structures. Therefore the trained DrugGPS model on the CrossDocked dataset also tends to generate many polycyclic structures as shown in Figure. \ref{case}. However, the fragment-based generation scheme used in DrugGPS enables us to flexibly generate the molecules we want. For example, if we want more monocyclic compounds, we can reduce polycyclic motifs in the motif library (e.g., increase the threshold $\tau$) and penalize merging rings at the generating stage (modified DrugGPS). In Figure. \ref{monocyclic}, we show the generated molecules with monocyclic structures by the modified DrugGPS. 

\textbf{Results on more datasets.} We understand comprehensive evaluations on more datasets could further improve our work. Here we further evaluate our method on experimentally determined protein-ligand complexes found in Binding MOAD \cite{hu2005binding}. Following \cite{schneuing2022structure}, Binding MOAD is filtered and split based on the proteins’ enzyme commission number, resulting in 40,354 protein-ligand pairs for training and 130 pairs for testing. In the following table, we compare DrugGPS with selected baselines including Pocket2Mol, FLAG, and DiffSBDD \cite{schneuing2022structure}. For a fair comparison, the hyperparameters are finetuned for Pocket2Mol and FLAG. We use the recommended hyperparameters for Binding MOAD dataset in the original paper \cite{schneuing2022structure} for DiffSBDD. We can observe that DrugGPS can also achieve competitive performance on the Binding MOAD dataset.

\begin{table*}[h]
\renewcommand{\arraystretch}{1.2}
\renewcommand\tabcolsep{2.9pt}
\centering
\scriptsize
\caption{Results on the Binding MOAD dataset.}
\vskip 0.15in
\begin{tabular}{c|ccccccccc}
\toprule
     \makecell[c]{Methods}& \makecell[c]{Vina Score\\(kcal/mol, ↓)} & \makecell[c]{High\\ Affinity(↑)} & \makecell[c]{QED (↑)} & SA (↑) & LogP & \makecell[c]{Lip. (↑)}  &\makecell[c]{Sim. Train (↓)} &\makecell[c]{Div. (↑)} &\makecell[c]{Time (↓)} \\ \midrule
     
    Testset&-8.103$\pm$2.26&- &0.602$\pm$0.15 &0.336$\pm$0.08&0.456$\pm$1.15&4.838$\pm$0.37&-&-&- \\
    
    DiffSBDD &\makecell[c]{-6.234$\pm$1.76} & \makecell[c]{0.127$\pm$0.11} & \makecell[c]{0.529$\pm$0.17}&\makecell[c]{0.324$\pm$0.10}&\makecell[c]{0.112$\pm$1.20}
    & \makecell[c]{4.847$\pm$0.33} & \makecell[c]{0.369$\pm$0.14} & \makecell[c]{0.717$\pm$0.09}&\makecell[c]{\textbf{463.0$\pm$171.4}} \\ 
    
    \makecell[c]{Pocket2Mol} &  \makecell[c]{-7.690$\pm$2.47} & \makecell[c]{0.358$\pm$0.22} & \makecell[c]{0.596$\pm$0.14} & \makecell[c]{\textbf{0.329$\pm$0.07}} & \makecell[c]{0.697$\pm$1.66}&
    \makecell[c]{4.750$\pm$0.28}&
    \makecell[c]{0.375$\pm$0.15}&
    \makecell[c]{\textbf{0.720$\pm$0.16}}&\makecell[c]{2618.0$\pm$1503.4}\\

    \makecell[c]{FLAG} &\makecell[c]{-7.724$\pm$2.09} & \makecell[c]{0.364$\pm$0.20} & \makecell[c]{0.605$\pm$0.16} & \makecell[c]{0.317$\pm$0.14} & \makecell[c]{0.719$\pm$1.43}&
    \makecell[c]{4.762$\pm$0.23}&
    \makecell[c]{0.382$\pm$0.16}&
    \makecell[c]{0.695$\pm$0.12}&\makecell[c]{1028.5$\pm$474.9}\\ 
    
    \makecell[c]{DrugGPS} & \makecell[c]{\textbf{-8.215$\pm$2.29}} & \makecell[c]{\textbf{0.522$\pm$0.21}} & \makecell[c]{\textbf{0.623$\pm$0.18}} & \makecell[c]{0.341$\pm$0.12}&\makecell[c]{0.560$\pm$1.41}&\makecell[c]{\textbf{4.859$\pm$0.24}}&\makecell[c]{\textbf{0.363$\pm$0.16}} &\makecell[c]{0.692$\pm$0.13}&\makecell[c]{1007.8$\pm$554.1}\\
\bottomrule
\end{tabular}
\vskip -0.1in
\label{main results sequence based}
\end{table*}

\end{document}